\documentclass[12pt]{article}

\renewcommand{\epsilon}{\varepsilon}

\newcommand{\dst}{\displaystyle}

\usepackage{amsmath}
\usepackage{amssymb}
\usepackage[ansinew]{inputenc}
\usepackage{graphicx}
\usepackage{epsfig}

\usepackage[authoryear]{natbib}
\newtheorem{theorem}{Theorem}[section]
\newtheorem{corollary}{Corollary}[section]
\newtheorem{lemma}{Lemma}[section]
\newtheorem{example}{Example}[section]
\newtheorem{remark}{Remark}[section]
\newtheorem{definition}{Definition}[section]

\def\IM{{\bf I\kern-.25em M}}

\def\3{\ss}

\newcommand{\bea}{\begin{eqnarray*}}
\newcommand{\eea}{\end{eqnarray*}}
\newcommand{\be}{\begin{eqnarray}}
\newcommand{\ee}{\end{eqnarray}}

\newcommand{\ba}{\begin{array}}
\newcommand{\ea}{\end{array}}
\def\3{\ss}

\newcommand{\ve}{\varepsilon}

\newcommand{\bS}{\mathbf{S}}

\newcommand{\btheta}{\theta}

\title{Optimal  designs in regression with  correlated errors}

\begin{document}

\author{Holger Dette  \\
Ruhr-Universit\"at Bochum \\
Fakult\"at f\"ur Mathematik \\
44780 Bochum \\
Germany
\and
Andrey Pepelyshev, Anatoly Zhigljavsky \\
School of Mathematics\\
 Cardiff  University\\
  Cardiff, CF24 4AG\\
  UK
}
\date{}
\maketitle

\begin{abstract}
This paper  discusses  the problem of determining optimal   designs for regression models, when the observations are dependent and 
taken on an interval.  A complete solution  of this challenging  
optimal design problem is given  for  a broad class of regression models and covariance kernels. \\
 We propose a class of estimators which are only slightly more complicated than the ordinary least-squares estimators.
We then demonstrate that we can design the experiments, such that asymptotically the new estimators achieve the same precision as the 
best linear unbiased estimator computed for the whole trajectory
of the process.  As a by-product we   derive explicit expressions for the BLUE in the continuous time model and analytic expressions for the optimal designs in a wide class of regression models.  We also demonstrate that for a finite number of observations the precision of the proposed procedure, which includes the estimator and design, is very close to the best achievable. The results are illustrated on a few numerical examples.
\end{abstract}






\section{Introduction} 
\label{sec1}
\def\theequation{1.\arabic{equation}}
\setcounter{equation}{0}

 Optimal design theory is a  classical field of mathematical statistics with  numerous applications in life sciences, physics and engineering. In many cases the use of optimal or efficient designs yields
to a reduction of costs  by a statistical inference with a minimal number of experiments without loosing any accuracy. Most work on optimal design theory
concentrates on experiments with independent observations. Under this assumption the field is very well  developed and a powerful methodology for the construction of optimal designs has been established
 [see for example the monograph of \cite{pukelsheim2006}]. While important and elegant results have been derived
in the case of independence,   there exist  numerous situations where correlation between different
observations is present and these classical  optimal designs are not applicable.

The theory of optimal design for correlated observations is much less
developed and explicit results are only available in rare circumstances.  The challenging difficulty  consists here in the fact that - in contrast
to the independent case -   correlations  yield  to    non-convex optimization  problems and classical tools of convex optimization theory are not applicable.
Some exact optimal designs for specific linear models have been studied
in \cite{detkunpep2008,KisStehlik2008,harstu2010}.
Because explicit solutions of  optimal design problems for correlated
observations are rarely available, several authors have proposed to determine
optimal designs based on asymptotic
arguments [see for example  \cite{sackylv1966,sackylv1968}, \cite{bickherz1979}, \cite{N1985a}, \cite{zhidetpep2010}],
where the references differ in the asymptotic arguments used to embed the discrete (non-convex) optimization problem in a continuous (or
approximate) one.
However, in contrast to the uncorrelated case, this approach does not simplify the problem substantially and
due to the lack of convexity the  resulting approximate  optimal design problems are still extremely difficult to solve.
As  a consequence,  optimal designs have mainly  been determined  analytically for the location  model
(in this case the optimization problems are in fact convex) and  for a few one-parameter linear models [see
\cite{bolnat1982}, \cite{N1985a}, Ch.\ 4, \cite{naether1985b},
 \cite{pazmue2001}   and  \cite{muepaz2003} among others]. Only recently, \cite{DetPZ2012} determined (asymptotic) optimal designs for least squares estimation in models with more parameters under the additional assumption that the regression functions are eigenfunctions of an integral operator associated with the covariance kernel of the error process. However, due to this assumption, the class of models for which approximate optimal designs can be determined explicitly is rather small.

The present paper provides a complete solution  of this challenging  
optimal design problem for  a broad class of regression models and covariance kernels. 
Roughly speaking, we determine (asymptotic) optimal designs for a
  slightly modified ordinary least squares estimator   (OLSE), such that the new estimate and the corresponding optimal design achieve the same accuracy as the best unbiased linear estimate (BLUE) with corresponding optimal designs.

To be more precise, consider a general regression observation scheme   given by
\be
\label{eq:model}
   y(t_j)=\theta^T f(t_j)+\varepsilon(t_j),~~\quad  j=1,\dots,N \, ,
     \ee
     where
     $     \mathbb{E}[\varepsilon(t_j)]=0, $ $K(t_i,t_j) = \mathbb{E}[\varepsilon(t_i)\varepsilon(t_j)]$ denotes the covariance between observations at the points $t_i$ and $t_j$. ($ i,j=1,\dots,N $),  $\theta=(\theta_1, \ldots, \theta_m)^T$ is a vector of unknown parameters, $f(t)=(f_1(t), \ldots, f_m(t))^T$ is a vector of linearly independent functions, the explanatory variables $t_1,\dots,t_N$  vary in a compact interval, say $[a,b]$.
Parallel to model \eqref{eq:model} we also consider its {\it continuous time} version
\be \label{mod1cont}
y(t) = \theta^Tf(t)+ \varepsilon(t) ~,~~ t \in [a,b],
\ee
where the full trajectory of the process $\{y(t)|t\in[a,b]\}$ can be observed and $\{\varepsilon(t)|t\in[a,b]\}$  is a centered Gaussian process with covariance kernel $K$, i.e. $K(s,t) = \mathbb{E}[\varepsilon (s) \varepsilon (t)]$. This kernel   is assumed to be continuous throughout this paper.

We pay much attention to the one-parameter case and develop a general method for solving the  optimal design problem
in model \eqref{mod1cont} explicitly  for the
 OLSE, perhaps slightly modified. The new estimate and the corresponding  optimal design achieve the minimal  variance among
 all linear estimates (obtained by the BLUE).  In particular, our approach allows to calculate this optimal variance explicitly.
As a by-product we also identify the BLUE in the continuous time model  \eqref{mod1cont}.
  Based on these asymptotic considerations, we consider the  finite sample case and suggest  designs for a  new  estimation procedure
  (which is very similar to OLSE) with an efficiency very close  to the best possible (obtained by the BLUE and
  the corresponding optimal design), for any number of observations.
 In doing this, we show how to implement the  optimal strategies from the continuous time model in practice and demonstrate that even for very small sample sizes the loss of efficiency with respect to the best strategies based on the use of BLUE with a  corresponding optimal design can be considered as  negligible.
  We would like to point out at this point that - even in the one-dimensional case - the problem of numerically calculating optimal designs for the BLUE for a fixed sample size is an extremely challenging one due to the lack of convexity of the optimization problem.

 In our approach, the importance of the one-parameter design problem is also related to the fact that the   optimal design problem  for    multi-parameter models can be reduced component-wisely  to  problems in the one-parameter models. This gives us a way to generate analytically constructed {\it universally
 optimal designs}  for a wide range of continuous time multi-parameter  models of the form \eqref{mod1cont}.
Our technique is based on the observation that for a finite number of observations  we can always emulate the BLUE
in model \eqref{eq:model} by a different linear estimator.
To achieve that theoretically    we assign signs to the support points of a discrete design and not only weights in the one-parameter models, but in the multi-parameter case we use matrix weights. We then
determine ``optimal'' signs and weights and 
 consider the weak convergence of these `designs'' and estimators as the sample size converges to infinity. Finally, 
we prove the (universal) optimality of the limits in the continuous time model  \eqref{mod1cont}.

Theoretically, we  construct a sequence of designs  for either the pure  or a  modified OLSE, say
$\hat \theta_N$, such   that its variance or covariance matrix satisfies $\mbox{Var}(\hat{\theta}_N) \to D^*$ as
the sample size $N$ converges to infinity, where  $D^*$ is
the variance (if $m=1$) or covariance matrix (if $m>1$) for the BLUE  in the continuous time model  \eqref{mod1cont}.
In other words,    $D^*$ is the smallest possible variance (or covariance matrix with respect to the Loewner ordering)
of any unbiased linear estimator and any design.
 This makes the designs  derived in this paper very competitive in applications against the designs
  proposed by \cite{sackylv1966} and optimal designs constructed numerically for the BLUE (using the Brimkulov-Krug-Savanov algorithm, for example).
We emphasize once again that due to non-convexity the numerical construction of optimal designs for the BLUE is
extremely  difficult.
An additional advantage of our approach is that we can  analytically compute the  BLUE with the corresponding
optimal  variance (covariance matrix) $D^*$  in the continuous time model  \eqref{mod1cont}
and therefore monitor the proximity of different approximations to the optimal variance $D^*$ obtained by the BLUE.


The methodology developed in this paper results in a non-standard estimation and optimal design theory and  consists in a delicate interplay between new linear estimators and   designs in the models \eqref{eq:model} and \eqref{mod1cont}. For this reason let us briefly introduce various estimators, which we will often refer to in the following discussion.
Consider the model \eqref{eq:model} and
suppose that  $N$ observations are taken  at experimental conditions $t_1,\ldots,t_N$.
For the   corresponding  vector of observations  $\mathbf{Y}=(y(t_1), \ldots, y(t_N))^T$,
a general weighted  least squares estimator (WLSE) of $\theta$ is defined by
  \be
\mathrm{WLSE:} &~~~&\widehat \btheta_{WLSE}  = (\mathbf{X}^T\mathbf{W}\mathbf{X})^{-1}\mathbf{X}^T\mathbf{W} \mathbf{Y},
\label{eq:WLSE}
\ee
where $\mathbf{X}=(f_i(t_j))^{i=1,\ldots,m}_{j=1,\ldots,N}$ is an $N\!\times\! m$ design matrix and
 $\mathbf{W}$ is some  $N \times N$ matrix such that $(\mathbf{X}^T\mathbf{W}\mathbf{X})^{-1}$ exists. For any such $\mathbf{W}$ the estimator \eqref{eq:WLSE} is obviously unbiased. 
The covariance matrix of the estimator \eqref{eq:WLSE}  is given by
\be
 \mathrm{Var} (\widehat\btheta_{WLSE}) &=~
 (\mathbf{X}^T \mathbf{W}\,\mathbf{X})^{-1} \mathbf{X}^T \mathbf{W}\,\mathbf{\Sigma}\mathbf{W}^T\, \mathbf{X} (\mathbf{X}^T \mathbf{W}^T\,\mathbf{X})^{-1}, \label{eq:var-wls}
\ee
where  $\mathbf{\Sigma}=(K(t_i,t_j))_{i,j=1,\ldots,N}$ is an $N\!\times\! N$ matrix of variances/covariances.
  For the standard WLSE the matrix  $\mathbf{W}$ is  symmetric non-negative definite; in this case  $\widehat \btheta_{WLSE} $ minimizes the weighted sum of squares   $SS_\mathbf{W}(\theta)= (\mathbf{Y}-\mathbf{X} \theta)^T \mathbf{W} (\mathbf{Y}-\mathbf{X} \theta)$ with respect to $\theta$.
Important particular cases of estimators of the form \eqref{eq:WLSE} are the
OLSE,
the best unbiased linear estimate (BLUE) and
the signed least squares estimate (SLSE):
 \be
\mathrm{OLSE:} &~~~&\widehat \btheta_{OLSE} = (\mathbf{X}^T\mathbf{X})^{-1}\mathbf{X}^T \mathbf{Y},
  \label{eq:OLS}\\
\mathrm{BLUE:} &~~~&\widehat \btheta_{BLUE}  = (\mathbf{X}^T\mathbf{\Sigma}^{-1}\mathbf{X})^{-1}\mathbf{X}^T\mathbf{\Sigma}^{-1} \mathbf{Y},
\label{eq:BLUE} \\
\mathrm{ SLSE:} &~~~&\widehat\btheta_{SLSE}=(\mathbf{X}^T \bS\, \mathbf{X})^{-1}\mathbf{X}^T \bS\, \mathbf{Y}.
 \label{eq:SLS}
\ee

Here
$\bS$ is an $N\!\times\! N$ diagonal matrix with entries $+1$ and $-1$ on the diagonal; note that if $\bS \neq \mathbf{I}_N$  then  SLSE is not a standard WLSE.
While the  use of BLUE and OLSE is standard, the SLSE is less common. It was introduced in \cite{bolnat1982} and further studied in
Chapter 5.3 of \cite{N1985a}.
In the content of the present paper, the SLSE will turn out to be very useful for constructing optimal designs for OLSE and the BLUE in the model \eqref{mod1cont} with one parameter, where the full trajectory can be observed. Another estimate of $\theta$, which is not a special case of the WLSE, will be introduced in Section \ref{sec:multi} and  used in the multi-parameter models.

The remaining structure of the paper is as follows.
In Section \ref{sec4} we derive   optimal designs
for continuous time one-parameter models and discuss how to implement the designs in practice.
In Section \ref{sec:multi} we extend
the results of Section \ref{sec4} to multi-parameter models.
In Appendix \ref{appA} we discuss transformations of regression models and associated designs,
which are a main tool in the proofs of our result but also of own interest.
In particular, we provide an extension of the famous Doob representation
for  Gaussian processes [see \cite{doob1949heuristic} and \cite{mehr1965certain}],
which turns out to be a   very important ingredient in proving
the design optimality results of Sections~\ref{sec4} and \ref{sec:multi}. Finally, in Appendix \ref{appB}  we
collect some auxiliary statements and  proofs for the main results of this paper.

\section{Optimal designs for
one-parameter  models } \label{sec4}
\def\theequation{2.\arabic{equation}}
\setcounter{equation}{0}

 In this section we concentrate on the one-parameter model
 \begin{equation} \label{onepar}
 y(t_j) = \theta f(t_j) + \varepsilon (t_j); \qquad j=1,\dots,N,
 \end{equation}
 on the interval $[a,b]$ and its continuous time analogue,  where
 $\mathbb{E}[\varepsilon(t)]=0$ and $\mathbb{E}[\varepsilon(t)\varepsilon(t')]=K(t,t')$.
Our approach uses some non-standard ideas and estimators in linear models and therefore we begin this section with a careful explanation of the logic of the material.

\begin{itemize}
  \item[]  {\it Sect. \ref{sec:slse}.} Under the  assumption that the design space is finite we show in Lemma \ref{lem:2-1} that by
  assigning  weights and signs to the observation points $\{ t_1,\dots,t_N\}$ we can construct a  WLSE which is equivalent to the BLUE.
  Then,    we derive in Corollary \ref{cor:2.1} an explicit form for the optimal weights
  for a broad class of covariance kernels, which are called  {\it triangular covariance kernels}.
  \item[] {\it Sect. \ref{sec22}.} We demonstrate in Theorem \ref{th:as-opt-des-tr} that the optimal designs derived in Sect. \ref{sec:slse}
  converge weakly to a signed measure, if the cardinality of the design space  converges to infinity.
  \item[] {\it Sect. \ref{sec23}.}  We consider model \eqref{onepar} under the assumption that the full trajectory of the process $\{y(t)|t \in [a,b] \} $  can be observed. For the specific case of Brownian motion,
  that is $K(t,t^\prime)=\min\{t , t^\prime\}$, we prove analytically the optimality
  of the signed measure derived in Theorem \ref{th:as-opt-des-tr}   for OLSE.
  Then, in Theorem \ref{th:op-des-tr} we establish optimality of the asymptotic measures from Theorem \ref{th:as-opt-des-tr} for general covariance kernels. As a by-product we also identify the BLUE in the continuous time model \eqref{mod1cont} (in the one-dimensional case).
  For this purpose, we introduce  a transformation which maps any regression model with
  a triangular covariance kernel into another model with different triangular kernels.
  These transformations allow us to reduce any optimization problem to the situation considered in Theorem \ref{th:opt-des-Wiener},
  which refers to the case of Brownian motion.
  The construction of this map is based on an extension of
  the celebrated Doob's representation which will be
  developed in Appendix \ref{appA}.
  \item[] {\it Sect. \ref{sec24}.} We provide some examples of asymptotic optimal measures for specific models.
\item[] {\it Sect. \ref{sec:pract-implem-1dim}.} We introduce a practical implementation of the asymptotic theory derived in the  previous sections.
For a finite sample size we construct WLSE with corresponding designs
which   can achieve very high efficiency compared to the  BLUE with corresponding optimal design.
It turns out that these estimators are slightly modified OLSE, where only observations
at the end-points obtain a weight (and in some cases also a sign).
    \item[] {\it Sect. \ref{sec26}.} We illustrate the new methodology in several examples.
    In particular, we give a comparison  with the best known procedures based on BLUE
    and show that the loss in precision for     the procedures derived in this paper is negligible
    with our procedures being much simpler and more robust than the procedures based on BLUE.
    \end{itemize}

\subsection{Optimal designs for SLSE on   a finite design space}
\label{sec:slse}

In this section, we suppose that the design space for model \eqref{onepar} is finite, say  ${\cal T}=\{t_1,\ldots,t_N\}$, and  demonstrate
that in this case the approximate optimal designs for the SLSE (\ref{eq:SLS}) can be found explicitly.
Since we consider the SLSE (\ref{eq:SLS}) rather than the OLSE (\ref{eq:OLS}), a generic approximate design  on the design space ${\cal T}=\{t_1,\ldots,t_N\}$
is an arbitrary discrete signed measure
 $ \xi = \{t_1,\ldots,t_N;w_1,\ldots,w_N\}$, where $w_i=s_i p_i$, $s_i \in \{-1,1\}$, $p_i \ge 0$ ($i=1, \dots, N$)  and \mbox{$\sum_{i=1}^N p_i=1$}.
We assume that the support
$t_1,\ldots,t_N$ of the design is fixed  but the weights $p_1,\ldots,p_N$ and signs $s_1,\ldots,s_N$,
or equivalently the signed weights $w_i$, will be chosen to minimize
the variance of the SLSE  (\ref{eq:SLS}). In view of \eqref{eq:var-wls}, this variance is given by
\be
\label{eq:D-SLSw}
D (\xi)= \sum^N_{i=1}\sum^N_{j=1} K(t_i,t_j) w_i w_j f(t_i)f(t_j)\Big/\Big(\sum^N_{i=1} w_i f^2(t_i)\Big)^2  .
\ee

Note that this expression coincides with the variance of the WLSE \eqref{mod1cont}, where the matrix $\mathbf{W}$ is defined  by $\mathbf{W}=\mbox{diag}(w_1,\dots,w_N)$.

We assume
that $f(t_i) \neq 0$ for all $i=1,\ldots,N$. If $f(t_j)=0$ for some $j$ then the point $t_j$ can be removed
from the design space $\mathcal{T}$ without changing the  SLSE estimator, its variance  and the corresponding value
$ D (\xi)$.
In  the above definition of the weights $w_i$, we have  $\sum^N_{i=1} |w_i|=\sum^N_{i=1} p_i=1$. Note, however,
 that
the value of the criterion \eqref{eq:D-SLSw} does not change
if we change all the weights from  $w_i$  to $ c w_i$ ($i=1, \ldots, N$) for arbitrary $c \neq 0$.

Despite the fact that the functional $D$ in \eqref{eq:D-SLSw} is not  convex as a function of $(w_1,\dots,w_N)$,
the problem of determining the optimal design can be easily solved by a simple application of the  Cauchy-Schwarz inequality.
The proof of the following lemma is given in Appendix  \ref{appB} [see also  Theorem 5.3 in  \cite{N1985a}, where
 this result was proved in  a  slightly different form].

\begin{lemma}
\label{lem:2-1}
Assume that the matrix $\mathbf{\Sigma}=(K(t_i,t_j))_{i,j=1,\ldots,N}$
is positive definite and \mbox{$f(t_i)\neq 0$} for all $i=1,\ldots,N$. Then
the optimal weights $w_1^*,\dots,w_N^*$ minimizing \eqref{eq:D-SLSw} subject to the constraint \mbox{$\sum^N_{i=1} |w_i|=1 $} are given by
\be
\label{eq:weights}
 w_i^*=c\frac{\mathbf{e}_i^T \mathbf{\Sigma}^{-1}\mathbf{f}}{f(t_i)}; \qquad i=1,\dots,N,
\ee
where $\mathbf{f}=(f(t_1),\ldots,f(t_N))^T$, $\mathbf{e}_i=(0,0,\ldots,0,1,0, \ldots,0)^T \in \mathbb{R}^N$ is the $i$-th unit vector, and
$$c=\Bigl(\sum^N_{i=1} | \mathbf{e}_i^T \mathbf{\Sigma}^{-1}\mathbf{f} / f(t_i)| \Bigr)^{-1}.$$
Moreover, for the design $\xi^*=\{t_1,\ldots,t_N;w_1^*,\ldots,w_N^*\}$ with weights \eqref{eq:weights} we have
$D (\xi^*)=D^*$, where $D^*=1/(\mathbf{f}^T\,\mathbf{\Sigma}^{-1}\mathbf{f})$, the variance of the BLUE defined
in \eqref{eq:BLUE} using all observations    $t_1,\ldots,t_N$.
\end{lemma}

Lemma \ref{lem:2-1} shows, in particular, that the pair \{SLS estimate, corresponding optimal design~$\xi^*$\} provides
an unbiased estimator with the best  possible variance  for the one-parameter model \eqref{onepar}.
This results in a WLSE \eqref{mod1cont} with $\mathbf{W}^*= \mbox{diag}(w^*_1,\dots,w^*_N)$ which is BLUE.  In other words, by a slight modification of the OLSE we are able to emulate the BLUE using the appropriate design or WLSE.

While the statement of Lemma \ref{lem:2-1} holds for arbitrary kernels,
we are able to  determine the optimal weights $w_i^*$ more explicitly  for a broad class, which are called {\it triangular kernels} and are of the form
\be
\label{eq:cov_tr0}
K(t,t')=u(t)v(t')\;\;\; {\rm for }\;\; t \leq t',
\ee
where $u(\cdot)$ and $v(\cdot)$ are  some  functions on the interval  
$[a,b]$. Note that the majority of covariance kernels considered in literature belong to this class, see for example \cite{N1985a,zhidetpep2010} or \cite{Harman}.
The following result is a direct consequence of Lemma~\ref{lem:lemma1} from Appendix~\ref{appB} .

\medskip

\begin{corollary}
\label{cor:2.1}
Assume that   the covariance kernel $K(\cdot,\cdot)$ has the form \eqref{eq:cov_tr0} so that the matrix $\mathbf{\Sigma}=(K(t_i,t_j))_{i,j=1,\ldots,N}$ is positive definite and
has the entries $ K(t_i,t_j)= u_i v_j$ for $i \leq j$, where for $k=1, \ldots, N$ we denote $u_k=u(t_k)$,  $v_k=v(t_k)$,  and  also  $f_k=f(t_k) $, $q_k=u_k/v_k$.
  If $f_1 \neq 0 \ (i=1,\dots,N)$,    the weights in \eqref{eq:weights}
    can be represented explicitly as follows:
\be \label{w10}
 w_1^* &=&\frac{c}{f_1}  \left( \tilde\sigma_{11} {f_1} + \tilde\sigma_{12} {f_2} \right) =
 \frac{c\, u_2}{f_1 v_1 v_2 (q_2-q_1)}
 \Bigl( \frac { f_{{1}}}{ u_{1}  } -  \frac{f_2}{u_2} \Bigr) \, , \\ \label{w20}
 w_N^* &=&\frac{c}{f_N}  \left( \tilde\sigma_{N,N} {f_N} + \tilde\sigma_{N-1,N} {f_{N-1}} \right)
 = \frac{c }{f_N v_N  (q_N-q_{N-1})}
 \Bigl( \frac { f_{N}}{ v_{N}  } -  \frac{f_{N-1}}{v_{N-1}} \Bigr)  \, , \\ \label{w30}
 w_i^*&=&\frac{c}{f_i}  \left(\tilde\sigma_{i,i} f_{i} + \tilde\sigma_{i-1,i} f_{i-1}  + \tilde\sigma_{i,i+1} f_{i+1}\right) \\ \nonumber
 &=& \frac{c}{f_i v_i}
 \Bigl(
 \frac{(q_{i+1}-q_{i-1}) f_{i}} {v_{i} (q_{i+1}-q_{i}) (q_{i}-q_{i-1})}
 -
 \frac{f_{i-1}} {v_{i-1} (q_{i}-q_{i-1})}
-
 \frac{f_{i+1}} {v_{i+1} (q_{i+1}-q_{i}) }
    \Bigr) \, ,
  \ee
  for $i=2, \ldots, N-1$.
  In formulas \eqref{w10},  \eqref{w20} and \eqref{w30}, the quantity $\tilde \sigma_{ij}$ denotes the element in the position $(i,j)$ of the matrix
  $\mathbf{\Sigma}^{-1}=(\tilde \sigma_{ij})_{i,j=1,\ldots,N}$.
\end{corollary}

\subsection{Weak convergence of  designs} \label{sec22}

 In this section,   we consider the asymptotic properties of  designs with weights \eqref{w10} - \eqref{w30}.
Recall that  the design space is  an interval, say $[a,b]$, and that we assume a
triangular  covariance function of  the form \eqref{eq:cov_tr0}.
According to the discussion of triangular covariance kernels provided in Section 4.1 of Appendix \ref{appA}, the functions $u(\cdot)$ and $v(\cdot)$ are continuous and strictly positive on the interval $(a,b)$ and the function
$
q(\cdot)=u(\cdot)/v(\cdot)
 $
 is positive, continuous and strictly increasing on  $(a,b)$. We also assume that the regression function $f$ in \eqref{onepar} is
continuous and strictly positive on the interval $(a,b)$. We define
the transformation
\begin{equation} \label{qtraf}
Q (t) = {q(t)-q(a) \over q(b) - q(a)}
\end{equation}
and note that the function $Q: [a,b] \to [0,1]$ is  increasing on the interval $[a,b]$ with $Q (a)=0$ and $Q (b)=1$, that is $Q(\cdot)$ is a
cumulative distribution function (c.d.f.). For  fixed $N$ and $i=1,\ldots, N$, define $ z_{i,N}=(i-\frac12)/N $ and the  design points
\be
\label{eq:des}
t_{i,N}=  Q^{-1} \left( z_{i,N} \right), \;\; i=1,\ldots, N\, .
\ee

\begin{theorem}
\label{th:as-opt-des-tr}
Consider the  optimal design problem for the model \eqref{onepar}, where the error process $\ve(t)$ has the covariance kernel $K(t,s)$
of the form
\eqref{eq:cov_tr0}.
Assume that  $u (\cdot)$,    $v (\cdot)$,  $f (\cdot)$ and $q (\cdot)$ are strictly positive, twice continuously differentiable functions on the interval $[a,b]$. Consider the sequence of signed measures
 $$
 \xi_N=\{t_{1,N}, \ldots, t_{N,N}; w_{1,N}, \ldots, w_{N,N} \},
 $$
 where the support points $t_{i,N} $ are defined in \eqref{eq:des} and the weights $w_{i,N} $ are assigned to these points
 according to the rule \eqref{eq:weights} of Lemma~\ref{lem:2-1}. Then the sequence of measures $\{\xi_N\}_{N \in \mathbb{N}}$ converges in distribution to a
 signed measure $\xi^*$, which
   has   masses
\be
\label{eq4A}
P_a
= \frac{  c }{ f(a)v^2(a) q^\prime (a)}  \Big[  \frac{f(a)u^\prime(a)}{u(a)} - f^\prime(a)  \Big]\, ,\;\;
P_b= c \cdot  \frac{   h^\prime(b)}{f(b) v(b) q^\prime (b)}
\ee
at the points  $a$ and~$b$, respectively,
and the signed density
\be
\label{eqpA}
  p(t) =
-\frac{c}{  f(t ) v( t )} \Big[ \frac{h^{\prime }(t)  }{q^{\prime } (t)}  \Big]^\prime\,
\ee
(that is, the Radon-Nikodym derivative of $\xi^*$ with respect to the Lebesque measure) on the interval $(a,b)$,
where the function $h(\cdot)$ is defined by $h(t)=f(t)/v(t)$.

\end{theorem}

The proof of Theorem \ref{th:as-opt-des-tr}  is   technically complicated and therefore   given in  Appendix \ref{appB}.
The constant $c \neq 0$ in \eqref{eq4A} and \eqref{eqpA} is arbitrary. If a normalization $| \xi^* | ([a,b])=1$ is required, then $c$ can be found from the normalizing condition
$$
\int^b_a \xi^* (dt)=|P_a|+|P_b|+\int_a^b |p(t)| dt=1\, .
$$

Throughout this paper we write the limiting designs of Theorem~\ref{th:as-opt-des-tr} in the form
\begin{equation} \label{optdes}
\xi^*(dt)=P_a\delta_a(dt)+P_b\delta_b(dt)+p(t)dt\, ,
\end{equation}
where $\delta_a(dt)$ and $\delta_b(dt)$ are the Dirac-measures concentrated at the points $a$ and $b$, respectively, and the function $p(\cdot )$
is defined by  \eqref{eqpA}. Note also that under the assumptions of Theorem~\ref{th:as-opt-des-tr}, the function $p(\cdot)$ is continuous on the interval $[a,b]$.
In the case of Brownian motion,  the limiting design of Theorem~\ref{th:as-opt-des-tr} is particularly simple.

\begin{example} \label{wienerexample} {\rm If the error process $\ve$ in model \eqref{onepar} is the  Brownian motion  on the interval $[a,b]$ with $0<a<b<\infty$,  then   $K(t,s) = \min (t,s)$ and hence
$u(t)=t$, $v(t)=1$,  $q(t)=t$. This implies that  the limiting design  of Theorem \ref{th:as-opt-des-tr}
is given by \eqref{optdes} with
\be \label{mass0a}
P_{a}=c \ \frac {f(a)-f^\prime (a) a}{a f(a)} \, ,\;\;
 P_{b}=c \ \frac {f^\prime (b)}{f(b)} \qquad {\rm{and}} \qquad  p (t)= - c \frac {f ^{\prime \prime}(t)}{f(t)}\,.
\ee
}
\end{example}

\subsection{Optimal designs and the BLUE} \label{sec23}

In this section we consider the continuous time model \eqref{mod1cont} in the case $m=1$ and  demonstrate that the limiting designs derived in Theorem \ref{th:as-opt-des-tr} are in fact optimal. A linear estimator for the parameter $\theta$  in model \eqref{mod1cont} is defined by $\hat \theta_\mu=\int^b_a y(t) \mu(dt)$, where $\mu$ is a signed measure on the interval $[a,b]$. Special cases include the OLSE and SLSE $\widetilde {\theta_\xi} = \int^b_a y(t) f(t) \xi (dt)   /   \int^b_a f^2(t) \xi (dt)$, where $\xi$ is a measure or a signed measure on the interval $[a,b]$, respectively. Note that $\hat \theta_\mu$ is unbiased if and only if
$\int^b_a f(t) \mu (dt)=1$ and $\widetilde {\theta_\xi}$ is unbiased by construction. The BLUE (in the continuous time model \eqref{mod1cont}) minimizes
$$
\Phi (\mu) = \mbox{Var} (\hat \theta_\mu) = \int^b_a \int^b_a K(x,y) \mu (dx) \mu (dy)
$$
in the class of all signed measures  $\mu$ satisfying $\int^b_a f(t) \mu (dt)=1$, and
\be \label{optBLUE}
D^* = \inf \{ \Phi(\mu) \mid \mu \ \ \mbox{signed measure on} \   [a,b] \}
\ee
denotes the best possible variance of all linear unbiased estimators in the continuous time model \eqref{mod1cont}.

Similarly, a  signed measure $\xi^*$ on the interval $[a,b]$ is called optimal  for least squares estimation in  the one-parameter model \eqref{mod1cont},
if it minimizes the functional
\be
  \label{eq:Phi}
 D (\xi) = \mbox{Var}(\widetilde{\theta_\xi}) =  {\int_{ a}^{ b} \int_{a}^{ b}  K(t,s)  f(t)  f(s)  \xi (dt)  \xi (d s) } \Big/ { \Big(\int_{ a}^{ b}  f^2(t)  \xi (dt)\Big)^2}\, ,
\ee
in   the set of all  signed measures $\xi $ on the interval $[a,b]$, such that $\int_{ a}^{ b}  f^2(t)  \xi (dt) \neq 0$.
In the case of a Brownian motion,  we are able to establish the optimality of the design of Example \ref{wienerexample}.
A proof of the following result is given in Appendix \ref{appB}.

\begin{theorem}
 \label{th:opt-des-Wiener}
 Let $\{\ve(t)\,|\,t \in [a,b]\}$ be a Brownian motion, so that $ K(t,t')=\min\{t,t'\}$, and $f$ be a positive, twice continuously differentiable function on the interval $[a,b] \subset \mathbb{R^+}$. Then
the signed measure $ \xi ^*$,  defined by \eqref{optdes} and \eqref{mass0a} with arbitrary $c \neq 0$, minimizes
the functional \eqref{eq:Phi}. The minimal value in \eqref{eq:Phi} is  obtained as 
$$
D(\xi^*)= \min_\xi D(\xi) =  \Big[\frac {f^2(a)}{a} + \int^b_a (f^\prime(t))^2 dt \Big]^{-1}.
$$
Moreover, the BLUE in model \eqref{mod1cont}  is given by $\hat \theta_{\mu^*}$, where $\mu^*(dt) = f(t) \xi^{**}(dt)$ and 
$\xi^{**}$ is the signed measure defined by \eqref{optdes} and \eqref{mass0a} with constant 
$
c^*=D(\xi^*)$. This further implies $ D^* =D(\xi^*)   = \Phi(\mu^*).$
\end{theorem}

Based on the design optimality established in Theorem \ref{th:opt-des-Wiener} for the special case of Brownian motion
and the  technique of transformation of regression models described in Appendix \ref{appA},
 we can establish the optimality of the asymptotic designs derived in
 Theorem \ref{th:as-opt-des-tr} for more general covariance kernels; see Appendix \ref{appB}  for the proof.

\begin{theorem}
\label{th:op-des-tr}
Under  the conditions of Theorem \ref{th:as-opt-des-tr}, the   optimal  design $\xi^*$ minimizing the functional \eqref{eq:Phi} is defined by the formulas
\eqref{eq4A}  -  \eqref{optdes} with arbitrary $c \neq 0$.
The minimal value in \eqref{eq:Phi} is obtained as 
\be
\label{Dstar1}
 D(\xi^*) = 
\Big[ \frac {\tilde{f}^2(q(a))}{q(a)} + \int^{q(b)}_{q(a)} (\tilde{f}^\prime(t))^2 dt \Big]^{-1} \, ,
\ee
where
 $
\tilde{f}(t)= f(q^{-1}(s))/v(q^{-1}(s))
$.
Moreover, the BLUE in model \eqref{mod1cont} is given by $\hat \theta_{\mu^*}$, where
 $\mu^*(dt)=   f(t) \xi^{**}(dt)$, $\xi^{**}$ is the signed measure defined in  \eqref{eq4A} - \eqref{optdes}
  with constant $c^* =D(\xi^*)$,
 and
 $D^* = \Phi (\mu^*) =  D(\xi^*).
 $

%
\end{theorem}

\subsection{Examples of  optimal designs} \label{sec24}

In this section,
 we provide the values of $P_a$, $P_b$ and the function $p(\cdot )$ in the general expression
\eqref{optdes}
for the optimal designs in a number of important special cases for the one-parameter continuous time model \eqref{mod1cont}, where
the design space is  ${\cal T}=[a,b]$. 
Specifically, optimal designs are given in Table \ref{tab:des4loc-model} for the location model,
in Table \ref{tab:des4lin-model} for the linear model,
in Table \ref{tab:des4quad-model} for a quadratic model
and in Table~\ref{tab:des4trig-model} for a trigonometric model.
The last named model  was especially chosen   to demonstrate the existence of   optimal designs with a density $p$
  which changes  sign in  the interval $(a,b)$.
  In the tables  several triangular covariance kernels are considered. The parameters of these covariance kernels satisfy the constraints $c_2>\pm c_1$, $\mp c_2\not\in[a,b]$, $\gamma>\omega$, $\lambda>0$. For the sake of a transparent presentation, we use the factor $c=1$ in all tables, but we emphasize once again that the optimal designs do not depend on the scaling factor. 
  
  As an  example, if $K(t,t^\prime) = e^{- \lambda|t-t^\prime|}$ for some $\lambda > 0$, we have from the last row of Table~\ref{tab:des4lin-model} that the   optimal design for the continuous time model $\{ \theta t + \varepsilon (t) | t \in [1,2]\}$ is 
$ \xi^*(dt)=
(\lambda-1) \delta_1 (dt) + (\lambda + \frac {1}{2}) \delta_2 (dt) + \lambda^2 dt
$, and  as a consequence,   $D^*= (\frac {5}{2} + \frac {1}{2  \lambda}+ \frac {7}{6} \lambda)^{-1}$.

\begin{table}[!hhh]
  \caption{\it{Optimal designs for the location model: $f(t)=1$, $t\in[a,b]$.}
  }
  \label{tab:des4loc-model}
  \smallskip
 \centering
 {\small
  \begin{tabular}{cc  ccc }
  \hline
  $u(t)$&$v(t)$&$P_a$&$P_b$&$p(t)$\\
  \hline
  any & $1$ & 1 & 0 & 0\\
  $c_1+t$ & $c_2\pm t$ & $\dst  \frac{1}{a+c_1}$& $\dst \frac{-1}{b\pm c_2}$ & $0$\\
  $t^\gamma$ & $t^\omega$ & $- \gamma a^{-\gamma-\omega}$ & $ \omega b^{-\gamma-\omega}$ & $ \gamma\omega t^{-1-\gamma-\omega}$\\
  $e^{\lambda t}$ & $e^{-\gamma t}$ &$ \lambda e^{a(\gamma-\lambda)}$ &$ \gamma e^{b(\gamma-\lambda)}$ &$ \lambda\gamma e^{t(\gamma-\lambda)}$\\
\hline
\end{tabular}
}
\end{table}

\begin{table}[!hhh]
  \caption{\it{Optimal designs  for the linear  regression model through the origin: $f(t)=t$, $t\in[a,b]$. }}
  \label{tab:des4lin-model}
  \smallskip
  \centering
  {\small
  \begin{tabular}{cc  ccc }
  \hline
  $u(t)$&$v(t)$&$P_a$&$P_b$&$p(t)$\\
  \hline
  $t$ & $1$ & 0 & 1 & 0\\
  $c_1+t$ & $c_2\pm t$ & $\dst  \frac{-c_1}{(a+c_1)a}$&$\dst  \frac{\pm c_2}{(b\pm c_2)b}$ & 0\\
  $t^\gamma$ & $t^\omega$ & $-\! (\gamma\!-\!1) a^{-\gamma-\omega}$ & $ (\omega\!-\!1) b^{-\gamma-\omega}$ &
  $ (1\!-\!\gamma)(1\!-\!\omega) t^{-1-\gamma-\omega}$\\
  $e^{\lambda t}$ & $1$ & $\dst  \frac{(a\lambda-1)e^{-a\lambda}}{a}$& $\dst  \frac{e^{-b\lambda}}{b}$& $\dst \frac{\lambda e^{-t\lambda}}{t}$\\
  $e^{\lambda t}$ & $e^{-\gamma t}$ &$\dst  \frac{a\lambda-1}{a}e^{a(\gamma-\lambda)}$&
  $\dst  \frac{b\gamma+1^{\rule{0mm}{3mm}}}{b}e^{b(\gamma-\lambda)}$&
  $\dst  \frac{\lambda\gamma t-\gamma+\lambda}{t}e^{t(\gamma-\lambda)}$\\
\hline
\end{tabular}
}
\end{table}

\begin{table}[!hhh]
  \caption{\it{Optimal designs  for the quadratic  regression model:  $f(t)=t^2+\nu$, $t\in[a,b]$.}}
  \label{tab:des4quad-model}
  \smallskip
  \centering
{\small
  \begin{tabular}{cc  ccc }
  \hline
  $u(t)$&$v(t)$&$P_a f(a)$&$P_b f(b)$&$p(t)f(t)$\\
  \hline
  $t$ & $1$ & $\dst  (a^2-\nu)/a$& $\dst -2 b$ & $\dst 2 $\\
  $c_1+t$ & $c_2\pm t$ & $\dst  \frac{(a^2-\nu+2ac_1)}{a+c_1}$&$\dst  \frac{\mp (b^2-\nu \pm 2bc_2)}{b\pm c_2} $
  &$\dst 2 $\\
  $t^\gamma$ & $t^\omega$ & ${\small  {((2\!-\!\gamma)a^{2}\!-\!\gamma\nu) a^{-\gamma-\omega}} }$&
  ${\small  {((\omega\!-\!2)b^{2}\!+\!\omega\nu) b^{-\gamma-\omega}} }$ &
  {\small{$  {((2\!-\!\omega\!)(2\!-\!\gamma)\!+\!\nu\gamma\omega) t^{1-\gamma-\omega}}$}}\\
  $e^{\lambda t}$ & $1$ & $\dst  {(2a-(a^2+\nu)\lambda)e^{-a\lambda}}$&$\dst -2{be^{-b\lambda}}$&$
 \dst {2 (1-t\lambda)e^{-t\lambda}}$\\
  $e^{\lambda t}$ & $e^{-\lambda t}$ &$\dst  {(2a-(a^2+\nu)\lambda)}$ & $\dst - {((b^2+\nu)\lambda+2b^{\rule{0mm}{3mm}} )}$ &
  $\dst  \left(  2-\lambda^2 (t^2+\nu) \right)$\\
\hline
\end{tabular}
}
\end{table}

\begin{table}[!hhh]
  \caption{\it{Optimal designs for the trigonometric  regression model:  $f(t)=1+\frac12 \sin(2\pi t)$, $t\in[1,2]$.}}
  \label{tab:des4trig-model}
  \smallskip
  \centering
{\small
  \begin{tabular}{cc  ccc }
  \hline
  $u(t)$&$v(t)$&$P_a$&$P_b$&$p(t)f(t)$\\
  \hline
  $t$ & $1$ & $\dst  (1-\pi)$& $\dst  \pi$ & $\dst 2 \pi^2\sin(2\pi t)$\\
  $c_1+t$ & $c_2\pm t$ & $\dst  \frac{1-\pi c_1-\pi}{c_1+1}$&$\dst \mp  \frac{1\mp \pi c_2 -2\pi}{ c_2\pm 2} $
  &$\dst 2 \pi^2\sin(2\pi t)$\\
  $t^2$ & $t$ & $\dst  (2-\pi)$&
  $\dst  (2\pi-1)/{8}$ &
  $\dst 2 t^{-4}{((\pi^2t^2\!-\!1)\sin(2\pi t)\!+\!\pi t \cos(2\pi t)\!-\!1)}{}$\\
  $e^{\lambda t}$ & $1$ & $\dst   (\lambda-\pi)e^{-\lambda}$&$\dst  \pi e^{-2\lambda}$&$
 \dst  (2\pi^2\sin(2\pi t)+\pi \lambda\cos(2\pi t))e^{-\lambda t}$\\
  $e^{\lambda t}$ & $e^{-\lambda t}$ & $\dst   (\lambda-\pi)$&$\dst  (\lambda+\pi) $&$
 \dst  ((2\pi^2+\lambda^2/2)\sin(2\pi t)+ \lambda^2)$\\
\hline
\end{tabular}
}
\end{table}

\subsection{Practical implementation: designs for finite sample size}
\label{sec:pract-implem-1dim}

In practice, efficient designs and corresponding estimators for the model \eqref{eq:model} have to be derived from the   optimal solutions in the continuous time model \eqref{mod1cont}, and in this section a procedure with a  good finite sample performance is proposed. Roughly speaking, it consists of a slight modification of the ordinary least squares estimator and a discretization of a continuous signed measure with the asymptotic optimal density in \eqref{eqpA} .

We assume that the experimenter can take $N\!+\!2$ observations with $N$ observations inside the interval $[a,b]$.
In principle, any probability measure on the interval can be approximated by an  $(N\!+\!2)$-point measure with weights $1/(N\!+\!2)$ and similarly
any finite signed  measure can be approximated by an $(N\!+\!2)$-point signed measure with equal weights (in absolute value). We hence could use a direct approximation of the optimal signed measures of the form \eqref{optdes} by a sequence of $(N\!+\!2)$-point signed measures with equal weights (in absolute value). For an increasing sample size this sequence will eventually converge to the optimal measure of Theorem~\ref{th:op-des-tr}.
However, this convergence will typically be very slow,
where we measure the speed of convergence by the differences between  the variances $D(\xi)$ of the corresponding estimates and the optimal value
$D^*$ defined in \eqref{Dstar1}.
The main difficulty  lies in the fact that a typical optimal measure has masses at 
the boundary points $a$ and $b$, in addition to some density on the interval  $(a,b)$. The convergence
of discrete measures with equal (in absolute value) weights to such a measure will be very slow, especially in view of the fact that
 in our approximating measures the points cannot be repeated.
Summarizing, approximation of the optimal signed measures  by measures with equal weights is possible but cannot be accurate for small $N$.  

In order to improve the rate of convergence we propose a slight modification of the ordinary least squares procedure. In particular, we propose     a WLSE with   weights at the points $a$ and $b$ (the end-points of the interval $[a,b]$), which correspond to the masses $P_a$ and $P_b$ of the asymptotic optimal design.
We thus only need to  approximate the continuous part of the optimal signed measure, which has a density on $(a,b)$,
by an $N$-point design with equal masses.
 To be precise, consider an optimal measure  of the form
 \eqref{optdes}. We assume that the density $p(\cdot )$ is not identically zero on the interval $(a,b)$ and
 choose the constant $c$ such that  \mbox{$\int_a^b |p(t)|dt =1$}.
Note that unless $p(\cdot )$ changes sign in $(a,b)$,  we can  choose $p(t)\geq 0$ for all $t \in (a,b)$.
Define
$\varphi(t) = |p(t)|$ for $ t \in (a,b)$ and denote by
$F(t)= \int_a^t \varphi(s) ds$  the  corresponding distribution function.
The $N$-point design we use as an $N$-point approximation to the measure with density $\varphi(t)$ is
$\hat{\xi}_{N}=\{ t_{1,N}, \ldots,t_{N,N}; 1/N, \ldots, 1/N\}  $, where $t_{i,N}= F^{-1}(z_{i,N})$ with $z_{i,N}= i/(N+1)$, $i=1,2, \ldots, N$.
 If $p(t)=0 $ on a sub-interval of $[a,b]$  and $ F^{-1}(z_{i,N})$ is not uniquely defined then  we choose the smallest element
from the set $ F^{-1}(z_{i,N})$ as $t_{i,N}$.
Finally, the design we suggest as an ($N\!+\!2$)-point approximation to the optimal measure in
 \eqref{optdes} is
 $$
\xi^*_{N\!+\!2}=P_{a}\delta_a+P_{b}\delta_b+P \bar \xi_{N},$$
where $P=1-|P_a|-|P_b|$,
 $\bar {\xi}_{N}=\{ t_{1,N}, \ldots,t_{N,N}; s_{1,N}/N, \ldots, s_{N,N}/N\}$ and
$
s_{i,N}= {\rm sign} (p(t_{i,N})),
$
$i=1, \ldots, N$. \\
The matrix $\mathbf{W}$, which corresponds to  the design $\xi_{N\!+\!2}$ and is  used in the corresponding  WLSE \eqref{eq:WLSE},
is a diagonal matrix  $\mathbf{W}_N = {\rm diag}(N P_{a}, s_{1,N}P, s_{2,N}P,
\ldots, s_{N,N}P, N P_{b})$ of size $(N\!+\!2)\!\times \!(N\!+\!2)$.
The set of $N\!+\!2$ design points, where the observations should be taken, is   given by
$\{a, t_{1,N}, t_{2,N}, \ldots, t_{N,N}, b\}$ and the resulting estimate is
defined by 
\be
\label{wlse}
   \widehat\btheta_{WLSE,N}   =
 (\mathbf{X}^T \mathbf{W}_N  \mathbf{X})^{-1} \mathbf{X}^T \mathbf{W}_N \mathbf{Y}.
\ee
It follows from \eqref{eq:var-wls},  \eqref{eq:Phi} and the discussion of the previous paragraph that
$$
\lim_{N \to \infty} \mathrm{Var} (\hat \theta_{\rm WLSE,N}) = \lim_{N \to \infty} D(\xi^*_{N\!+\!2}) = D^*,
$$
where $D^*$ is defined in \eqref{optBLUE}.

\subsection{Some numerical results}
\label{sec26}

Consider the regression model \eqref{onepar} with $f(t)=t^2+1$, $t\in[1,2]$, where the error process is given by the  Brownian motion.
The optimal design for this model can be obtained from Table~\ref{tab:des4quad-model}, and we have $P_a=0$, $P_b=-0.55$, $P=0.45$ and $p(t)=1.38/(t^2+1)$.
By computing the quantiles from the c.d.f. corresponding to $p $ we can easily obtain
support points of $(N\!+\!2)$-point designs. For example, $\mathrm{supp}\,(\xi^*_4)=\{1,1.24,1.56,2\}$,
$\mathrm{supp}\,(\xi_5^*)=\{1,1.18, 1.39, 1.65,2\}$ and
$\mathrm{supp}\,(\xi_6^*)=\{1,1.14, 1.30, 1.49, 1.71,2\}$.

\begin{figure}[!hhh]
\centering
  \includegraphics[width=80mm]{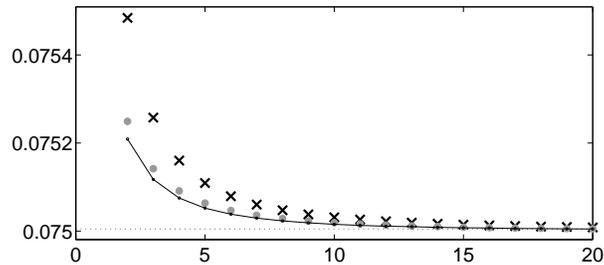}
\caption{
\it{The variance of the WLSE defined in \eqref{wlse} for the proposed $(N\!+\!2)$-point designs $\xi^*_{N+2}$  (crosses),
  of the BLUE for the proposed $(N\!+\!2)$-point designs (grey circles)
and of the BLUE with corresponding optimal $(N\!+\!2)$-point designs (line). The error process in model \eqref{onepar} is given by the Brownian motion and the regression function is $f(t)=t^2+1$,
$t\in[1,2]$.   }}
\label{fig:bluewlsvar-quad}
\end{figure}

In Figure \ref{fig:bluewlsvar-quad} we display the variance of various linear  unbiased estimators for different sample sizes. We observe
that the variance of the WLSE defined by \eqref{wlse} for the proposed $(N\!+\!2)$-point design  $\xi^*_{N+2}$
is slightly larger than the variance of the  BLUE for the proposed $(N\!+\!2)$-point design,
which is very  close to the variance of the BLUE with corresponding optimal $(N\!+\!2)$-point design.
The calculation of these designs is complicated and has been performed numerically by the Nelder-Mead algorithm in MATLAB. We also note that due to the non-convexity of the optimization problem it is not clear that the algorithm finds the optimal design. However, by Theorem \ref{th:opt-des-Wiener} and \ref{th:op-des-tr} we determined the optimal value \eqref{optBLUE}, which is $D^* \simeq 0.075004$.
This means that for the proposed designs WLSE has almost the same precision as BLUE.

\begin{figure}[!hhh]
\centering
   \includegraphics[width=80mm]{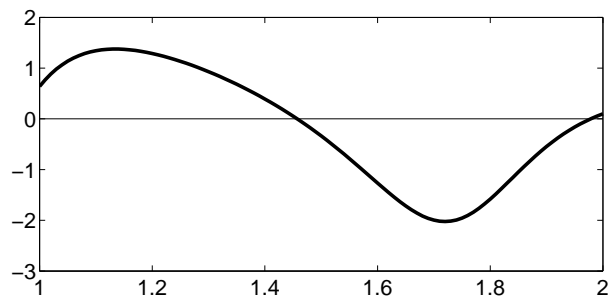}
\caption{\it{The density of the  optimal design for continuous time  model \eqref{onepar} with regression function  $f(t)=1+0.5\sin(2 \pi t)$, $t\in[1,2]$,
and   covariance kernel of the form \eqref{eq:cov_tr0} with $u(t)=t^2$ and $v(t)=t$.}}
\label{fig:optdens-trig}
\end{figure}

In our second example we     compare the proposed optimal designs with the designs from \cite{sackylv1966},
which are constructed for the BLUE.
For this purpose we consider the model \eqref{onepar} with regression function $f(t)=1+0.5\sin(2 \pi t)$, $t\in[1,2]$,
and triangular covariance kernel of the form \eqref{eq:cov_tr0} with $u(t)=t^2$ and $v(t)=t$.
The optimal design in the continuous time model can be obtained from Table \ref{tab:des4trig-model} and its density is depicted in Figure \ref{fig:optdens-trig}.

\begin{figure}[!hhh]
\centering
   \includegraphics[width=80mm]{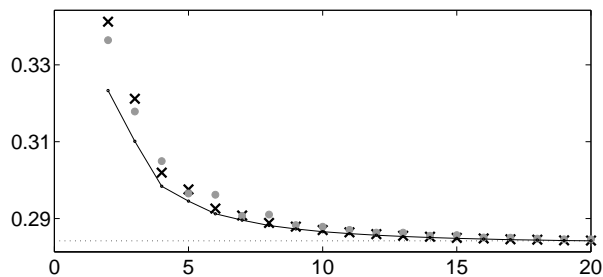}
\caption{\it{The variance of BLUE for the proposed $(N\!+\!2)$-point designs (grey circles),
the $(N\!+\!2)$-point designs from \cite{sackylv1966} (crosses)
and the BLUE with corresponding optimal $(N+2)$-point designs (line) for the model $f(t)=1+0.5\sin(2 \pi t)$, $t\in[1,2],$
and the covariance kernel with $u(t)=t^2$ and $v(t)=t$; $N=2,\ldots,20$. 
}}
\label{fig:bluevar-trig}
\end{figure}

By computing quantiles using this optimal design, we obtain
that the 4-point design $\xi^*_4$ is supported at points 1, 1.27, 1.68 and 2.
For $\xi^*_4$, the variance of the BLUE is $ \simeq 0.6129$.
Using the optimal density from \cite{sackylv1966}, we obtain the 4-point design $\xi^{SY}_4$ supported at 1, 1.25, 1.63 and 2.
For $\xi^{SY}_4$, the variance of the BLUE is $\simeq 0.6200$.
For $N=2,3,\ldots,20$, the variances of the BLUE for the  proposed $(N+2)$-point designs,
the $(N+ 2)$-point designs from \cite{sackylv1966} and the optimal $(N+2)$-point designs for the BLUE
are depicted in Figure~\ref{fig:bluevar-trig}.
We observe that for $N=2,3,4$ the new designs yield a smaller variance of the BLUE, while for $N=5$ the design of \cite{sackylv1966}
shows a better performance. In all other cases the results for both designs are very similar.
In particular, for  $N\ge 6$ the  variances from the optimal $(N\!+\!2)$-point
designs proposed in this paper and  in the paper of \cite{sackylv1966}
are only slightly worse than the variances of the BLUE with corresponding best $(N\!+\!2)$-point designs
(which is  computed by direct optimization).

\section{Multi-parameter models}
\label{sec:multi}
\def\theequation{3.\arabic{equation}}
\setcounter{equation}{0}

In this section we discuss optimal design problems for models with more than one parameter.
{The structure of this section is somewhat similar to the structure of Section \ref{sec4}.}
In Section \ref{sec31}  we introduce a new class of linear estimators of the parameters in model \eqref{eq:WLSE},
  which we call {\it matrix-weighted estimators} (MWE) and
    show in Lemma \ref{lem:blue} that for some special choices of the matrix weights the MWE can   always emulate the BLUE.
In Section \ref{sec32}  matrix-weighted designs associated with the MWE are defined.
 Then, for the case of triangular kernels,  in Corollary \ref{th:mult-as-opt-des-tr} we derive the  asymptotic forms for the sequence of  designs that are associated
 with the version of  the MWE which emulates the BLUE.
In Section  \ref{sec33}  we prove optimality
  of the asymptotic matrix-weighted measure derived in Corollary \ref{th:mult-as-opt-des-tr} in the continuous time model \eqref{mod1cont}
  (see  Theorem \ref{th:mult-op-des-tr}),
  while  some examples of asymptotically optimal measures are provided in Section  \ref{sec34}. Finally,
the  practical implementation of the asymptotic measures is discussed in Section \ref{sec35} and numerical examples are provided
in Section  \ref{sec36}. \\
The proofs of many statements in this section use the results of Section \ref{sec4}. This is   possible as there is a lot of freedom in choosing the form of the MWE to emulate the BLUE and we choose a special form  which could be considered as component-wise SLSE. Correspondingly, the resulting matrix-weighted designs (including the asymptotic ones) become combinations of designs for one-parameter models.


\subsection{Matrix-weighted estimators and designs} \label{sec31}

Consider the regression model \eqref{eq:model} and assume that $N$ observations at points $t_j$ ($j=1,\ldots,N$) have been made.
Let $\mathbf{O}_j$ be an $m\times m$ matrix    associated with the observation point $t_j$; $j=1,\ldots,N.$ Recall the definition of the design matrix $\mathbf{X}=(f_i(t_j))^{i=1,\ldots,m}_{j=1,\ldots,N}$
and the definition of $\mathbf{Y}=(y(t_1),\ldots,y(t_N))^T.$
We introduce the $m \times N$ matrix
$
 \mathbf{C}=  (\mathbf{O}_1f(t_1),\ldots,\mathbf{O}_Nf(t_N) )\, ,
$
whose $j$-th column is $\mathbf{O}_jf(t_j)$.
 Assuming that the $m  \times m $ matrix
\be
\label{eq:M}
\mathbf{M}= \mathbf{C}\mathbf{X}=\sum_{j=1}^N \mathbf{O}_j f(t_j)f^T(t_j)
\ee
is non-singular  we define  the   linear estimator
\be
 \widehat\btheta_{MWE}=(\mathbf{C} \mathbf{X})^{-1} \mathbf{C}\mathbf{Y}
 \label{eq:MWE}
\ee
for the vector $\theta$ in model \eqref{eq:model}. We  call this estimator the  {\it matrix-weighted estimator} (MWE), because each column of the matrix $\mathbf{X}$ is multiplied by a matrix weight.
It is easy to see that for any $\mathbf{C}$ the MWE $\widehat\btheta_{MWE}$ is unbiased  and
its covariance matrix is given by
\be
\label{eq:VarMWE}
 \mbox{Var}(\widehat\btheta_{MWE})=\mathbf{M}^{-1}\mathbf{C}\mathbf{\Sigma}\mathbf{C}^T (\mathbf{M}^{-1})^T,
\ee
where $\mathbf{\Sigma}=(K(t_i,t_j))_{i,j=1,\ldots,N}$ is the $N\!\times\! N$ matrix of covariances of the errors.
Note that the matrix $\mathbf{M}$ defined in \eqref{eq:M} generalizes the standard information matrix $\mathbf{X^TX}$
and that $\mathbf{M}$ is not necessarily a symmetric matrix.
The following result shows that  different  matrices $\mathbf{O}_1,\ldots,\mathbf{O}_N$ may yield the same matrix-weighted estimator $\widehat\btheta_{MWE}$. Its proof is obvious and therefore omitted.

\begin{lemma}
 \label{lem:LC=CforMWE}
 Consider the regression model \eqref{eq:model} and assume that the matrix $\mathbf{M}$ defined in \eqref{eq:M} is non-singular.
 Then the estimator $\widehat\btheta_{MWE}$ defined in \eqref{eq:MWE} coincides with the estimator
 $\widehat\btheta_{MWE,\Lambda}=(\mathbf{C}_\Lambda \mathbf{X})^{-1} \mathbf{C}_\Lambda\mathbf{Y}\, ,$
 where $\mathbf{C}_\Lambda=\Lambda \mathbf{C} $ and $\Lambda$ is an arbitrary non-singular $m\times m$ matrix.
 \end{lemma}

The estimator $\widehat\btheta_{MWE,\Lambda}$ introduced in Lemma~\ref{lem:LC=CforMWE} is the MWE defined by the matrix weights
$\Lambda \mathbf{O}_1, \ldots, \Lambda \mathbf{O}_N$. Lemma~\ref{lem:LC=CforMWE} implies that the $ \widehat\btheta_{MWE}$ is exactly the same for any set of matrices $\{\Lambda \mathbf{O}_1, \ldots, \Lambda \mathbf{O}_N\}$ as long as $\Lambda $ is non-singular.
In the asymptotic considerations below it will be convenient to interpret the combination of the set of experimental conditions $\{t_1, \ldots, t_N\}$ and
the set of corresponding matrices $\{\mathbf{O}_1, \ldots, \mathbf{O}_N\}$ in the MWE as an $N$-point matrix-weighted design.

\medskip

\begin{definition} \label{def3.1}
{Any combination of $N$ points $\{t_1, \ldots, t_N\}$ and  $m \times m$ matrices $\{\mathbf{O}_1, \ldots, \mathbf{O}_N\}$
will be called $N$-point matrix-weighted design and denoted by
\be
\label{mwd}
\xi_N = \big\{t_1, \ldots, t_N;  \frac1N  \mathbf{O}_1, \ldots, \frac1N  \mathbf{O}_N\big\}\,.
\ee
The covariance matrix $\mathbf{D}(\xi_N)$ of a matrix-weighted design $\xi_N$ is defined as the covariance matrix ${\rm Var}(\widehat\btheta_{MWE})$   in \eqref{eq:VarMWE}
  of the corresponding estimate $\widehat\btheta_{MWE}$.}
 \end{definition}

\medskip

The estimator $\widehat\btheta_{MWE}$ is not necessarily a least-squares type estimator; that is,
it may not be representable in the form \eqref{eq:WLSE}
for some $N\times N$ weight matrix $\mathbf{W}$
and hence  there may be no associated weighted sum of squares which is minimized by the MWE.
However, for any given $\mathbf{W}$, we can always find matrices $\mathbf{O}_j$ such that
\be
 \mathbf{C}=\mathbf{X}^T\mathbf{W}
\label{eq:C=XW}
\ee
and therefore achieve $\widehat\btheta_{MWE}=\widehat\btheta_{WLSE}$.
The following result gives a constructive solution to the matrix equations \eqref{eq:C=XW}.

\begin{lemma}
\label{lem:omega=}
Assume that $f_1(t)\neq 0$ for all $t\in [a,b]$.
Define $\mathbf{O}_j=\omega_je_1^T$,
\be
\omega_j=\frac{1}{f_1(t_j)}(\mathbf{X}^T \mathbf{W})_j \in \mathbb{R}^m\, ,
\label{eq:omega=XW/f}
\ee
where $e_1=(1,0,\ldots,0)^T \in \mathbb{R}^m$ is the first unit vector
and $(\mathbf{X}^T \mathbf{W})_j$ denotes the $j$-th column of the $m \times N$ matrix $\mathbf{X}^T \mathbf{W}$.
Then the corresponding matrix-weighted estimator satisfies $\widehat\btheta_{MWE}=\widehat\btheta_{WLSE}$.
\end{lemma}

\textbf{Proof.}
The matrix equation \eqref{eq:C=XW} can be written as $N$ vector equations
\be
 \mathbf{O}_j f(t_j)=(\mathbf{X}^T \mathbf{W})_j; \qquad j=1,\dots,N,
  \label{eq:Of=XW}
\ee
with respect to the matrices $\mathbf{O}_j$.
Assume that $\mathbf{O}_j=\omega_je_1^T$ for some  $\omega_j \in \mathbb{R}^m$.
Then
\bea
 \mathbf{O}_j f(t_j)=\omega_je_1^Tf(t_j)=\omega_j f_1(t_j)
\eea
and   equation  \eqref{eq:Of=XW}  has the unique solutions \eqref{eq:omega=XW/f}.
\hfill$\Box$

\medskip

\medskip

The form $\mathbf{O}_j=\omega_je_1^T$ for the matrices $\mathbf{O}_j$ considered in Lemma~\ref{lem:omega=}
means that the matrix $\mathbf{O}_j$ has the vector $\omega_j$ as its first column while
all other entries in this matrix are zero. We shall refer to this form as the {\it one-column form}.
We can choose other forms for the matrices $\mathbf{O}_j$,  but then we  would require different, somewhat stronger, assumptions regarding the vector  $f(t)$.
For example, if $f(t)\neq (0,\ldots,0)^T$ for all $t \in [a,b]$, then we can always choose diagonal matrices $\mathbf{O}_j$ to satisfy \eqref{eq:C=XW} (see Lemma~\ref{lem:omegaD}   below).

The following choices for  $\mathbf{O}_j$ ensure coincidence of  $\widehat\btheta_{MWE}$  with the three  popular estimators defined in the Introduction.

\medskip

If $\mathbf{O}_j=\mathbf{I}_m$ for all $j$, then
$\widehat\btheta_{MWE}=\widehat \btheta_{OLSE}$.

If $\mathbf{O}_j=s_j\mathbf{I}_m$ for all $j$, then
$\widehat\btheta_{MWE}=\widehat\btheta_{SLSE}$.

If $\mathbf{W}=\mathbf{\Sigma}^{-1}$ and $\mathbf{O}_j=\omega_je_1^T$ with $\omega_j= (\mathbf{X}^T \mathbf{\Sigma}^{-1})_j/{f_1(t_j)}$,
then $\widehat\btheta_{MWE}=\widehat \btheta_{BLUE}$.

\medskip

We shall call any MWE $\widehat\btheta_{MWE}$ optimal if it coincides with the BLUE. In view of the  importance of the last case, the corresponding result  is summarized in the following lemma.

\begin{lemma}
\label{lem:blue}
Consider the   regression model  \eqref{eq:model} and let $f_1(t) \neq 0$ for all $t \in [a,b]$.
For a given
set of $N$ observation points $\{ t_1, \ldots, t_N\}$
the MWE  $\widehat\btheta_{MWE}$ defines a BLUE if
$\mathbf{O}_j=\omega_j^*e_1^T$ with $\omega_j^*= (\mathbf{X}^T \mathbf{\Sigma}^{-1})_j/{f_1(t_j)}$.
\end{lemma}

If the covariance kernel of the error process has triangular form \eqref{eq:cov_tr0} then we can
derive the explicit form for the optimal MWE. The result follows  by a direct application of Lemma~\ref{lem:lemma1}.

\begin{lemma}
\label{lem:blue1}

Assume that   the covariance kernel $K(\cdot,\cdot)$ has the form \eqref{eq:cov_tr0} and that 
the matrix $\mathbf{\Sigma}=(K(t_i,t_j))_{i,j=1,\ldots,N}$ is positive definite with entries $ K(t_i,t_j)= u_i v_j$ for $i \leq j$, where for $k=1, \ldots, N$ we denote $u_k=u(t_k)$,  $v_k=v(t_k)$ and   $q_k=u_k/v_k$.
Then we have
the following representation for the optimal vectors   $\omega_j^*= (\mathbf{X}^T \mathbf{\Sigma}^{-1})_j/{f_1(t_j)}  \in \mathbb{R}^m$ introduced in Lemma~\ref{lem:blue}:
\be \label{w1}
 \omega_1^* &=&\frac{c}{f_1(t_1)}  \left( \tilde\sigma_{11} f(t_1) + \tilde\sigma_{12} {f(t_2)} \right) =
 \frac{c\, u_2}{f_1(t_1) v_1 v_2 (q_2-q_1)}
 \left( \frac { f(t_{1})}{ u_{1}  } -  \frac{f(t_2)}{u_2} \right) \, , \\ \label{w2}
 \omega_N^* &=&\frac{c}{f_1(t_N)}  \left( \tilde\sigma_{N,N} {f(t_N)} + \tilde\sigma_{N-1,N} {f(t_{N-1})} \right) \\ \nonumber
 &=& \frac{c }{f_1(t_N) v_N  (q_N-q_{N-1})}
 \left( \frac { f(t_{N})}{ v_{N}  } -  \frac{f(t_{N-1})}{v_{N-1}} \right)  \, , \\
 \omega_i^* &=&\frac{c}{f_1(t_i)}  \left(\tilde\sigma_{i,i} f(t_i) + \tilde\sigma_{i-1,i} f(t_{i-1})  + \tilde\sigma_{i,i+1} f(t_{i+1})\right) \label{w3} \\ \nonumber
 &=& \frac{c}{f_1(t_i) v_i}
 \left(
 \frac{(q_{i+1}-q_{i-1}) f(t_{i})} {v_{i} (q_{i+1}-q_{i}) (q_{i}-q_{i-1})}
 -
 \frac{f(t_{i-1})} {v_{i-1} (q_{i}-q_{i-1})}
-
 \frac{f(t_{i+1})} {v_{i+1} (q_{i+1}-q_{i}) }
    \right) \, ,
  \ee
  for $i=2, \ldots, N-1$. Here in formulas \eqref{w1}, \eqref{w2} and \eqref{w3}
  $\tilde \sigma_{ij}$ denote the elements of the matrix
  $\mathbf{\Sigma}^{-1}=(\tilde \sigma_{ij})_{i,j=1,\ldots,N}$.
\end{lemma}

The following provides a result similar to  Lemmas \ref{lem:omega=} and \ref{lem:blue}
 in the case where the matrices $\mathbf{O}_j$ are diagonal. An extension of Lemma~\ref{lem:blue1} to the matrices $\mathbf{O}_j$ of the diagonal form is straightforward and omitted for the sake of brevity.
 \begin{lemma}
\label{lem:omegaD}
Consider the  regression model  \eqref{eq:model} and let $f_k(t) \neq 0$ for all $t \in [a,b]$ and all $k=1,\ldots,m$.
For each $j=1,\ldots,N$, define the diagonal matrix $\mathbf{O}_j$ by its diagonal elements
\bea
(\mathbf{O}_j)_{k,k}=\frac{1}{f_k(t_j)}(\mathbf{X}^T \mathbf{W})_{k,j}; \qquad k=1,\ldots,m\, ,
\eea
where $(\mathbf{X}^T \mathbf{W})_{k,j}$ denotes the $(k,j)$-th element of the matrix $\mathbf{X}^T \mathbf{W}$.
Then $\widehat\btheta_{MWE}=\widehat\btheta_{WLSE}$.

If additionally $\mathbf{W}= \mathbf{\Sigma}^{-1}$ so that $(\mathbf{O}_j)_{k,k}=(\mathbf{X}^T \mathbf{\Sigma}^{-1})_{k,j}/f_k(t_j)$, then
$\widehat\btheta_{MWE}=\widehat\btheta_{BLUE}$.
\end{lemma}

\subsection{Weak convergence of matrix-weighted designs}  \label{sec32}

Let $Q: [a,b] \to [0,1]$ be an    increasing function on the interval $[a,b]$ with $Q (a)=0$ and $Q (b)=1$ so that $Q(\cdot)$ is a
c.d.f.
For a fixed $N$ and $j=1,\ldots, N$, define    the  points $t_{1,N}, \dots, t_{N,N}$ by \eqref{eq:des}.
Suppose that with each $t \in [a,b]$ we can associate an $m\times m$ matrix $\mathbf{O}(t)$ and consider an
$N$-point matrix-weighted design $\xi_N$ of the form
\eqref{mwd} with $t_j=t_{j,N}$ and $\mathbf{O}_j=\mathbf{O}(t_{j,N})$.
In view of \eqref{eq:M} and \eqref{eq:VarMWE} this design  has the covariance matrix
$$
\mathbf{D}(\xi_N) = \mathbf{M}^{-1}(\xi_N)  \mathbf{B}(\xi_N)   \left(\mathbf{M}^{-1}(\xi_N) \right)^T \, ,
$$
where the matrices $M(\xi_N)$ and $B(\xi_N)$ are defined by
\bea
\mathbf{M}(\xi_N) & = & \frac {1}{N}   \sum_{j=1}^N \mathbf{O}(t_{j,n}) f(t_{j,n})f^T(t_{j,n}), \\
\mathbf{B}(\xi_N) & = & \frac {1}{N^2} \sum_{i=1}^N  \sum_{j=1}^N K(t_{i,n},t_{j,n}) \mathbf{O}(t_{i,n}) f(t_{i,n})f^T(t_{j,n}) \mathbf{O}^T(t_{j,n})    .
\eea

In addition to the  sequence of matrix-weighted designs  $\xi_N$ consider  the sequence of uniform distributions on the set
$
  \{t_{1,N}, \ldots, t_{N,N} \}.
 $
As $N \to \infty$,  this sequence  converges weakly to the design (probability measure) $\zeta$ on the interval $[a,b]$ with distribution function $Q$. 
This implies
 $$
\lim_{N \to \infty} \mathbf{M}(\xi_N) = \mathbf{M}(\xi) = \int_a^b \mathbf{O}(t) f(t)f^T(t)  \zeta( d t)
$$
$$
\lim_{N \to \infty} \mathbf{B}(\xi_N) = \mathbf{B}(\xi)= \int_a^b \int_a^b K(t,s) \mathbf{O}(t) f(s) f^T(t) \mathbf{O}^T(s)  \zeta(d t)  \zeta(d s)\, ,
$$
and
\be
\label{D_Matrix}
\lim_{N \to \infty} \mathbf{D}(\xi_N) = \mathbf{D}(\xi)=\mathbf{M}^{-1}(\xi)  \mathbf{B}(\xi)   \left(\mathbf{M}^{-1}(\xi) \right)^T
\ee
under the assumptions that the vector-valued function $f$, the matrix-valued function $\mathbf{O}$,   the kernel $K$ are continuous on the interval $[a,b]$ and the generalized information  matrix $\mathbf{M}(\xi)$ are non-singular.
Moreover,
 the sequence of estimators
 \eqref{eq:MWE} converges (almost surely as $N \to \infty$) to
\be
\label{MWEasymp}
\widehat\btheta_{MWE, \infty}=\mathbf{M}^{-1}(\xi) \int_a^b \mathbf{O}(t) f(t) y(t)  \zeta(dt)\, ,
\ee
where $\{ y(t)\,|\,t \in [a,b] \}$ is the stochastic process in the continuous time   model \eqref{mod1cont}.
Bearing these limiting expressions in mind we  say that   the sequence of matrix-weighted designs $\xi_N$ defined by
\eqref{mwd}
converges to the limiting matrix-weighted design
$\xi(dt)=\mathbf{O}(t) \zeta(dt)$ as $N\to \infty$.
This relation justifies the notation $\mathbf{M}(\xi), \mathbf{B}(\xi)$ and $\mathbf{D}(\xi)$ of the previous paragraph.

The (optimal) limiting matrix-weighted designs which  will be constructed below will have a   similar structure as the signed measures  in   \eqref{optdes}. They will assign matrix weights
$\mathbf{O}_a$ and $\mathbf{O}_b$ to
the end-points of the interval $[a,b]$ and a `matrix density' $\mathbf{O}(t)$ to the points $t \in (a,b)$; that is, these designs will have the form
\begin{equation} \label{optdesM}
\xi(dt)=\mathbf{O}_a\delta_a(dt)+\mathbf{O}_b\delta_b(dt)+\mathbf{O}(t)dt\, .
\end{equation}

In view of \eqref{MWEasymp}, the MWE in the continuous time model \eqref{mod1cont} associated with any design of the form \eqref{optdesM} can be written as
\be
\label{MWEasymp2}
\widehat\btheta_{MWE} (\xi)=\mathbf{M}^{-1}(\xi) \Big[ \mathbf{O}_a f(a) y(a) + \mathbf{O}_b f(b) y(b) +  \int_a^b \mathbf{O}(t) f(t) y(t) dt   \Big]  \, ,
\ee
where
$
\label{MWEasymp3}
\mathbf{M}(\xi)=  \mathbf{O}_a f(a) f^T(a)  + \mathbf{O}_b f(b) f^T(b) +  \int_a^b \mathbf{O}(t) f(t) f^T(t) dt
$.
In the   particular case   associated with Lemma \ref{lem:blue1}, we have the following structure of the matrices $\mathbf{O}_a$ and $\mathbf{O}_b$ and the matrix function $\mathbf{O}(t)$ in \eqref{optdesM}:
\begin{equation} \label{optdesM0}
\mathbf{O}_a=\omega_a e_1^T,\;\mathbf{O}_b=\omega_b e_1^T,\;\, \mathbf{O}(t)=\omega(t) e_1^T\, \;{\rm for\;} t \in (a,b),
\end{equation}
where $\omega_a$ and $\omega_b $ are some $m$-dimensional vectors and $\omega(t) \in \mathbb{R}^m$ is some vector-valued function defined on the interval $(a,b)$. Note that
$\omega(t)$ does not have to approach   $\omega_a$ and $\omega_b $ as $t\to a$ and $t \to b$, respectively.

  When the sequence of matrix-weighted designs is defined by the  formulas of Lemma~\ref{lem:blue}   we can compute the limiting matrix-weighted design. The proof follows by similar arguments as given in the proof of Theorem \ref{th:as-opt-des-tr}  and is therefore omitted.

\begin{corollary}
\label{th:mult-as-opt-des-tr}
Consider    model \eqref{eq:model}, where the error process $\{ \ve(t)|\,t \in [a,b]\}$ has a covariance kernel $K$ of the form
\eqref{eq:cov_tr0}.
Assume that  $u (\cdot)$,    $v (\cdot)$, $q (\cdot)$
 are strictly positive, twice continuously differentiable functions on the interval $[a,b]$ and that the vector-valued function  $f (\cdot)$  is
  twice continuously differentiable with  $f_1(t) \neq 0 $ for all $t \in [a,b]$.
  Consider the matrix-weighted design    $\xi_N$ of the form  \eqref{mwd},
 where the support points $t_j=t_{j,N} $ are generated by \eqref{eq:des} and the matrix weights $\mathbf{O}_j=\mathbf{O}_{j,N} $
 are defined in Lemma~\ref{lem:blue}.
 The sequence $\{\xi_N\}_{N \in \mathbb{N}}$  converges (in the sense defined above in the previous paragraph) to a
  matrix-weighted design $\xi$ defined by  \eqref{optdesM} and  \eqref{optdesM0} with
 \be
\label{eq4}
\omega_a
\!=\! \frac{  c }{ f_1(a)v^2(a) q^\prime (a)}  \left[  \frac{f(a)u^\prime(a)}{u(a)} - f^\prime(a)  \right] ,\;
\omega_b\!=\!  \frac{ c\;  h^\prime(b)}{f_1(b) v(b) q^\prime (b)},\;  \omega(t) \!= \!
\frac{-c}{  f_1(t ) v( t )} \left[ \frac{h^{\prime }(t)  }{q^{\prime } (t)}  \right]^\prime\;\;\;\;
\ee
where $h(t)=f(t)/v(t)$ and the constant $c \neq 0$ is arbitrary.
\end{corollary}


In Corollary \ref{th:mult-as-opt-des-tr}, the  one-column representation of the matrices $\mathbf{O}_j$ is used.
The following statement  contains a similar  result  for the case where the matrices $\mathbf{O}_j$  are diagonal.

\begin{corollary}
\label{rem:diag-form}
Let the conditions of Corollary \ref{th:mult-as-opt-des-tr} hold and assume additionally that
    $f_k(t) \neq 0 $ for all $t \in [a,b]$ and all $k=1,\ldots,m$.
  Consider the matrix-weighted design    $\xi_N$ of the form  \eqref{mwd},
 where the support points $t_j=t_{j,N} $ are generated by \eqref{eq:des} and the matrices  $\mathbf{O}_j=\mathbf{O}_{j,N} $
 are defined in  Lemma~\ref{lem:omegaD} with diagonal elements  given by
$(\mathbf{O}_j)_{k,k}=(\mathbf{X}^T \mathbf{\Sigma}^{-1})_{k,j}/f_k(t_j)$.
 Then the sequence $\{\xi_N\}_{N \in \mathbb{N}}$  converges to
  the optimal matrix-weighted design $\xi^*$ of the form \eqref{optdesM},
  where the diagonal elements of the matrices $\mathbf{O}_a=\mathrm{diag}(\mathbf{O}_{a,11},\ldots,\mathbf{O}_{a,mm})$,
  $\mathbf{O}_b=\mathrm{diag}(\mathbf{O}_{b,11},\ldots,\mathbf{O}_{b,mm})$ and
  $\mathbf{O}(t)=\mathrm{diag}(\mathbf{O}_{11}(t),\ldots,\mathbf{O}_{mm}(t))$ are given by
$$
\mathbf{O}_{a,jj}
\!=\! \frac{  c }{ f_j(a)v^2(a) q^\prime (a)} \! \left[  \frac{f_j(a)u^\prime(a)}{u(a)} \!-\! f^\prime_j(a)  \right] ,\;
\mathbf{O}_{b,jj}\!=\!  \frac{ c\;  h^\prime_j(b)}{f_j(b) v(b) q^\prime (b)},\;
\mathbf{O}_{jj}(t) \!= \!
\frac{-c}{  f_j(t ) v( t )} \!\left[ \frac{h^{\prime }_j(t)  }{q^{\prime } (t)}  \right]^\prime  \;\;\;\;\;\;
$$
respectively, $h_j(t)=f_j(t)/v(t)$, $j=1,\ldots,m$ and the constant $c \neq 0$ is arbitrary.
\end{corollary}

\subsection{Optimal designs and best linear estimators }  \label{sec33}


In this section we consider again the continuous time model \eqref{mod1cont}, where the full
trajectory of the process $\{y(t) |~ t \in [a,b]\}$ can be observed. We start recalling
some known facts concerning best linear unbiased estimation. For details we refer the interested reader to
the work of  \cite{grenander1950} or  Section  2.2 in  \cite{N1985a}.
Any linear estimator of $\theta$ can be written in the form of the  integral
\be
\label{estSt}
\widehat{\theta}_\mu= \int_a^b y(t)  \mu (dt),
\ee
where $ \mu (t)= ( \mu_1(t),\ldots,  \mu_m(t))^T$ is a vector of signed measures  on  the interval  $[a,b]$.
For given $\mu$, the estimator
$\widehat{\theta}_\mu$ is unbiased if and only if $\int_a^b f(t)  \mu^T (dt) = \mathbf{I}_m $, where
$\mathbf{I}_m$ denotes the $m$-dimensional identity matrix.
Theorem 2.3 in  \cite{N1985a} states that the estimator $\widehat{\theta}_{\mu^*}$ is BLUE if and only if
$\int_a^b f(t) { \mu^*}^T (dt) = \mathbf{I}_m $ and the identity 
$$
\int_a^b K(u,v)\mu^* (dv)  = \mathbf{A} f(u)  
$$
holds  for all $u \in [a,b]$,
where $\mathbf{A}$ is some $m \times m$ matrix.   The matrix $\mathbf{A}$   is uniquely defined and coincides
 with the matrix
\begin{equation} \label{optmat}
\mathbf{D}^* = \mbox{Var}( \widehat{\theta}_{\mu^*})  = \inf \Big\{ \int_a^b\int_a^b K(u,v)\mu  (du)  {\mu  }^T(dv)  ~\Big| ~
 \mu \mbox{ vector of signed measures} \Big\} ~.
\end{equation}
The Gauss-Markov theorem further implies that $\mathbf{D}^*\leq \mbox{Var}(\widehat\btheta)$, where $\widehat\btheta$ is any other
linear unbiased estimator of
$\theta.$

\begin{definition} \label{def3.2} A matrix-weighted design $\xi^*$ is called optimal if $\mathbf{D}(\xi^*)=\mathbf{D}^*$,
where  $\mathbf{D}(\xi)$ is defined in  \eqref{D_Matrix} and  $\mathbf{D}^* $ is
defined in \eqref{optmat}.
\end{definition}

\medskip

The designs we consider have the form \eqref{optdesM} and the corresponding  MWE are expressed by \eqref{MWEasymp2}. The estimator \eqref{MWEasymp2} can be expressed
in the form \eqref{estSt}, that is  $\widehat\btheta_{MWE} (\xi)= \hat \theta_\mu$ with
$$
\mu (dt)= \mathbf{M}^{-1}(\xi) \bigl[ \mathbf{O}_a f(a) \delta_a (dt) + \mathbf{O}_b f(b) \delta_b (dt) +  \mathbf{O}(t) f(t)  dt   \bigr]\, .
$$

The estimators defined in \eqref{MWEasymp2} are always unbiased and
the following result provides the matrix-weighted optimal design and the   BLUE in the continuous time model \eqref{mod1cont}.
The proof follows by similar arguments as given in the proof of Theorem \ref{th:opt-des-Wiener} and \ref{th:op-des-tr} and is therefore omitted.

%
%
%
%
%
%

%

\begin{theorem}
\label{th:mult-op-des-tr}
Let $K(t,s)$ be a covariance kernel of the form \eqref{eq:cov_tr0} and
the vector-function  $f (\cdot) $  be
  twice continuously differentiable with  $f_1(t) \neq 0 $ for all $t \in [a,b]$.
Under  the assumptions of Corollary \ref{th:mult-as-opt-des-tr}
the   matrix-weighted design $\xi^*$ defined  by the formulas \eqref{optdesM} and
\eqref{eq4} with $c=1$
is optimal in the sense of Definition~3.2. Moreover, if
$$
\mu^*(dt) = \mathbf{M}^{-1}(\xi^*) \bigl[ \omega_a e_1^T  f(a) \delta_a (dt) +\omega_b e_1^T  f(b) \delta_b (dt) +  \omega  (t)  e_1^T f(t)  dt   \bigr]\, ,
$$
then $\hat  \theta_{\mu^*}$  defines the BLUE in  model \eqref{mod1cont}. Additionally, we have
\bea
 \mathbf{D}(\xi^*)=\mathbf{D}^*= \int_a^b \int_a^b K(s,t) \mu^*(ds) \mu^*(dt)  =
 \mathbf{M}^{-1}(\xi^*),
\eea
where the matrix $\mathbf{M}(\xi^*)$ is given by
\bea
 \mathbf{M}(\xi^*)= \frac{\tilde f(q(a))\tilde f^T(q(a))}{q(a)}+\int_{q(a)}^{q(b)} \tilde f'(s)\tilde f'^T(s)ds
\eea
and $\tilde f(s)=f(q^{-1}(s))/v(q^{-1}(s))$.
\end{theorem}

In Theorem \ref{th:mult-op-des-tr} we have used the one-column representation for the matrices $\mathbf{O}(t)$.
Similar arguments establish the optimality of the matrix-weighted designs $\xi^*$ defined in Corollary \ref{rem:diag-form}
where the diagonal representation for the matrices $\mathbf{O}(t)$ is used. The details are omitted for the sake of brevity.
\noindent

\subsection{Examples of  optimal matrix-weighted designs}  \label{sec34}

Consider the polynomial regression model with $f(t)=(1,t,t^2,\ldots,t^{m-1})^T$, $t\in[a,b]$ and
the covariance kernel of the Brownian motion $K(t,s)=\min(t,s)$. For the construction of matrix-weighted designs  we use matrices $\mathbf{O}(t)$ in the one-column and diagonal representations.

For the one-column representation   we have from Corollary \ref{th:mult-as-opt-des-tr} and Theorem \ref{th:mult-op-des-tr} that
the optimal matrix weighted  design has masses $\mathbf{O}_a=\omega_a e_1^T$ and
$\mathbf{O}_b=\omega(b)e_1^T$ at points $a$ and $b$, respectively,
and the density $\mathbf{O}(t)=\omega(t)e_1^T$. Here the vectors $\omega_a, \omega_b$ and $\omega(t)$ are given by
\bea
\omega_a&=&(1/a,0,-a,\ldots,(2-m)a^{m-2})^T,\\
\omega_b&=&(0,1,2b,3b^2,\ldots,(m-1)b^{m-2})^T,\\
\omega(t)&=&(0,0,-2,-3\cdot2t,\ldots,-(m-1)(m-2)t^{m-3})^T,~t\in(a,b),
\eea
respectively.
For the diagonal representation   we have  from Corollary \ref{rem:diag-form}  (and an analogue of Theorem \ref{th:mult-op-des-tr}) that
the optimal  matrix weighted   design has masses $\mathbf{O}_a$ and $\mathbf{O}_b$ at points $a$ and $b$, respectively,
and the density $\mathbf{O}(t)$,
where
\bea
\mathbf{O}_a&=&\mathrm{diag}(1/a,0,-1/a,\ldots,(2-m)/a),\\
\mathbf{O}_b&=&\mathrm{diag}(0,1/b,2/b,\ldots,(m-1)/b),\\
\mathbf{O}(t)&=&\mathrm{diag}(0,0,-2/t^2,\ldots,-(m-1)(m-2)/t^2),~t\in(a,b).
\eea
Note that in this case all non-vanishing diagonal elements of the matrix $\mathbf{O}(t)$ are proportional to  the function $1/t^2$.
According to Lemma \ref{lem:LC=CforMWE},
we can use $\Lambda \mathbf{O}(t)$ instead of $\mathbf{O}(t)$, for any non-singular $m\times m$ matrix $\Lambda$.
By taking the matrix $$\Lambda=\mathrm{diag}(1,1,-1/2,\ldots,-1/[(m-1)(m-2)]),$$
we obtain
\bea
\Lambda \mathbf{O}(a)\;&=&\mathrm{diag}(1/a,0,1/(2a),\ldots,1/[(m-1)a]),\\
\Lambda \mathbf{O}(b)\;&=&\mathrm{diag}(0,1/b,-1/b,\ldots,-1/[(m-2)b]),\\
\Lambda \mathbf{O}(t)&=&\mathrm{diag}(0,0,1/t^2,\ldots,1/t^2),~t\in(a,b).
\eea


As another example, consider the polynomial regression model with $f(t)=(1,t,t^2,\ldots,t^{m-1})^T$, $t\in[a,b]$, and
the triangular covariance kernel of the function \eqref{eq:cov_tr0} with $u(t)=t^\gamma$ and $v(t)=t^{\omega}$.

For the diagonal representation   we have from Corollary \ref{rem:diag-form} that
the optimal  matrix weighted design has masses $\mathbf{O}_a$ and $\mathbf{O}_b$ at points $a$ and $b$, respectively,
and the density $\mathbf{O}(t)$,
where
\bea
\mathbf{O}_a&=&a^{-\gamma-\omega}\mathrm{diag}(-\gamma ,1-\gamma,2-\gamma,\ldots,m-1-\gamma),\\
\mathbf{O}_b&=&b^{-\gamma-\omega}\mathrm{diag}(\omega ,\omega-1,\omega-2b,\ldots,\omega+1-m),\\
\mathbf{O}(t)&=&t^{-1-\gamma-\omega} \mathrm{diag}(\tau_1, \tau_2, \ldots, \tau_m),~t\in(a,b),
\eea
with
$ \tau_i= (i-1-\gamma)(i-1-\omega)$, $i=1, \ldots, m$.
If we further use $\Lambda=\mathrm{diag}(1/\tau_1, 1/\tau_2, \ldots, 1/\tau_m)$ then  we obtain
$\Lambda \mathbf{O}(t)=t^{-1-\gamma-\omega}  \mathrm{diag}(1, 1, \ldots, 1) ,~t\in(a,b)$; that is, all components of the matrix $\Lambda \mathbf{O}(t)$ have exactly the same density.

\subsection{Practical implementation }  \label{sec35}

 Here we only consider the diagonal representation of the matrices $\mathbf{O}_a$, $\mathbf{O}_b$ and $\mathbf{O}(t)$; the case of one-column representation of the matrices $\mathbf{O}$ can be treated  similarly.
We  assign matrix weights $\mathbf{O}_a$ and $\mathbf{O}_b$ to the boundary points $a$ and $b$ and
use an $N$-point approximation to an absolutely continuous  probability measure on $(a,b)$ with some density $\varphi(t) $.
The density $\varphi(t) $  is defined to be either the uniform density on $(a,b)$ (if nonzero elements of different components of  $\mathbf{O}(t)$
are not proportional to each other) or 
 $\varphi(t)=c |\mathbf{O}_{l,l}(t)|$ for some  $l \in \{1,\dots,m\}$ (if nonzero elements of different components of  $\mathbf{O}(t)$
are proportional to each other), where  $c$ is the normalization constant and $l$ is such that 
the density $\varphi(t)$ is not identically zero on  the interval  $(a,b)$. Denote by
$F(t)= \int_a^t \varphi(s) ds$  the  corresponding c.d.f.
For given $N$, we calculate  an $N$-point approximation $\{ t_{1,N}, \ldots,t_{N,N}; 1/N, \ldots, 1/N\}  $,
where $t_{i,N}= F^{-1}(z_{i,N})$ with $z_{i,N}= i/(N+1)$, $i=1,2, \ldots, N$, to the probability measure with density $\varphi(t)$.

To each point $t_{j,N}$ we assign  a vector of weights $s_j=(s_{j,N,1}, \ldots, s_{j,N,m})^T$
such that $s_{j,N,k} \in \{-1,0,1\}$ ($k=1, \ldots, m$). The values $s_{j,N,k}=\mathrm{sign}(\mathbf{O}_{k,k}(t_j)) =\pm 1$
correspond to the sign of the point $t_{j,N}$  in the estimation of  $\theta_k$ exactly as in the procedure for one-parameter models described in Section~\ref{sec:pract-implem-1dim}.
Some of the values
$s_{j,N,k}  $ could be 0. If $s_{j,N,k}=0  $  for some $k$ then
the point $t_{j,N}$ is not used for the estimation of  $\theta_k$.
By assigning zero weight to a point $t_{j,N}$ in the $k$-th
estimation direction, we perform a   thinning of the sample of points $t_{1,N},\ldots,  t_{N,N}$ in $k$-th direction 
and thus achieve a required density in the each estimation direction.
This is a deterministic version of the well-known `rejection method'
widely used to generate samples from various probability distributions.
If the nonzero  components of the matrix weight  $\mathbf{O}(t)$
are proportional to each other then for these components $s_{j,N,k}=1  $ for all $j$ and $N$.

The resulting estimator $\hat{\theta}$ has the form \eqref{eq:MWE}
where
\bea
 \mathbf{C}&=&\Big(N\mathbf{O}_af(a),\mathbf{S}_1\mathbf{P} f(t_1)\ldots,\mathbf{S}_N\mathbf{P} f(t_N),N\mathbf{O}_bf(b)\Big),\\
 \mathbf{S}_j&=&\mathrm{diag}(s_{j,N,1},\ldots,s_{j,N,m})\in\mathbb{R}^{m\times m}
\eea
and $\mathbf{P}$ is the diagonal $m\times m$ matrix whose diagonal elements are given by
$$
 \mathbf{P}_{k,k}=\frac{N}{\sum_{j=1}^N|s_{j,N,k}|}\int_a^b \mathbf{O}_{k,k}(t)dt.
$$

 If nonzero elements of different components of the  matrix weight    $\mathbf{O}(t)$
are proportional to each other (as was the case in the  examples of Section~\ref{sec34}) then the $(N\!+\!2)$-point approximations to the limiting design are very similar
to the approximations in the one-parameter case considered in Section~\ref{sec:pract-implem-1dim}; their accuracy is also very high.
 Otherwise, when the diagonal elements  of  $\mathbf{O}(t)$ are possibly non-proportional, the accuracy of approximations will depend on the degree of non-homogeneity  of components of the  matrix weight   $\mathbf{O}(t)$.

\subsection{Some numerical results}  \label{sec36}

\begin{figure}[!hhh]
\centering
  \includegraphics[width=80mm]{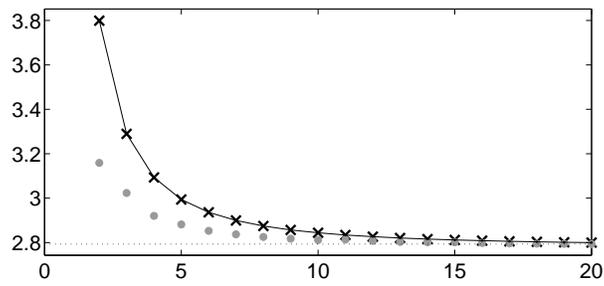}
\caption{
\it{The $D$-optimality criterion of the covariance matrix of the MWE for the proposed $(N\!+\!2)$-point designs  (crosses),
 of the covariance matrix of the BLUE for the proposed $(N\!+\!2)$-point designs (line)
and of the covariance matrix of the BLUE with corresponding $D$-optimal $(N\!+\!2)$-point designs (grey circles). The error process in model \eqref{eq:model} is the  Brownian motion and the vector of regression functions is given by $f(t)=(1,t,t^2,t^3)$,
$t\in[1,2]$.  }}
\label{fig:bluemwevar-cubic}
\end{figure}

For comparison of competing matrix weighted designs for multiparameter models it is convenient to consider
a functional of  the covariance matrix. Exemplarily we investigate in this Section  the classical $D$-optimality criterion  defined as
$
 \Psi(\mathbf{D}(\xi))=(\det \mathbf{D}(\xi))^{1/m}
$
which has to be minimized. \\
As an example where all nonzero elements of  the matrix $\mathbf{O}(t)$
are proportional to each other,
let us consider the cubic regression model with $f(t)=(1,t,t^2,t^3)^T$
and the Brownian motion error process.
The optimal value in the continuous time model \eqref{mod1cont} is 
$\Psi(\mathbf{D}^*)\simeq 2.7927$.
In Figure \ref{fig:bluemwevar-cubic} we display the $D$-criterion of the covariance matrices of
the MWE and the BLUE for the proposed $(N\!+\!2)$-point designs and
the covariance matrix of the BLUE  with corresponding optimal $(N\!+\!2)$-point designs.
We can see that  the $D$-efficiency of the proposed matrix-weighted design is very high, 
even for small $N$.

The second example of this section considers a situation where nonzero elements of the matrix $\mathbf{O}(t)$
are not proportional to each other.  For this purpose we   consider the model \eqref{onepar} with $f(t)=(1,t,t^2)^T$, $t\in[1,2]$ and
 covariance kernel $K(t,t')=e^{-|t-t'|}$  with $u(t)=e^t$ and $v(t)=e^{-t}$.
Using the diagonal representation, we obtain for the optimal matrix-weighted designs
\bea
\mathbf{O}_a=\mathrm{diag}(1,0,-1),\;\;
\mathbf{O}_b=\mathrm{diag}(1,1.5 ,2 ),\;\;
\mathbf{O}(t)=\mathrm{diag}(1,1,1-2/t^2),~t\in(1,2).
\eea
The optimal value in the continuous time model \eqref{mod1cont} is given by
$\Psi(\mathbf{D}^*)\simeq 1.6779$.
Since some diagonal elements of $\mathbf{O}(t)$ are constant functions,
we take the support points of the design $\xi_{N+2}$ to be equidistant: $t_{i,N}=i/(N+1)$ for $i=1,\ldots,N$.
Then we have $s_{j,N,k}=1$ for all $j=1,\ldots,N$ and $k=1,2$.
However, some elements of $(s_{1,N,3},\ldots,s_{N,N,3})$ should be zero
because $\mathbf{O}_{3,3}(t)$ is not proportional to $\mathbf{O}_{1,1}(t)$.
For example, for $N=10$ the vector of signs  $(s_{1,N,3},\ldots,s_{N,N,3})$ is 
$
(-1, -1, 0, 0, 0, 1, 0, 0, 1, 0)
$
and for $N=30$ it is\\
$
 (-1, 0, -1, -1, 0, -1, 0, 0, -1, 0, 0, 0, 0, 0, 0, 0, 0, 1, 0, 0, 0, 1, 0, 0, 1, 0, 0, 1, 0, 1)\, .
$

\begin{figure}[!hhh]
\centering
 \includegraphics[width=80mm]{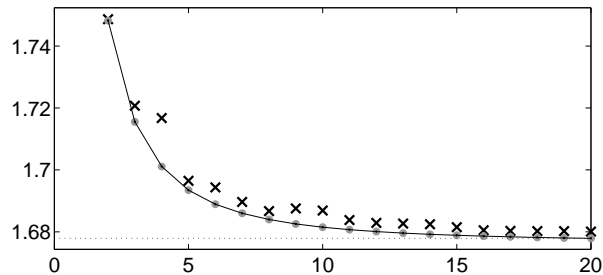}
\caption{
\it{The $D$-optimality criterion of the covariance matrix of the MWE for the proposed $(N\!+\!2)$-point designs  (crosses),
 of the covariance matrix of the BLUE for the proposed $(N\!+\!2)$-point designs (line)
and  of the BLUE with corresponding $D$-optimal $(N\!+\!2)$-point designs (grey circles). The covariance kernel in model \eqref{eq:model} is $K(t,t')=e^{-|t-t'|}$ and the vector of regression functions is $f(t)=(1,t,t^2)$,
$t\in[1,2]$.  }}
\label{fig:bluemwevar-quad-exp}
\end{figure}

In Figure \ref{fig:bluemwevar-quad-exp} we depict the $D$-optimality criterion of the covariance matrices for various estimators.
We observe that in this example for all $N$ the $D$-optimality criterion of the covariance matrices of the MWE is slightly larger
than  the $D$-optimality criterion
of the covariance matrices of the BLUE.
However, we can also see that the proposed $(N\!+\!2)$-point designs are
very efficient compared to the BLUE with corresponding $D$-optimal $(N\!+\!2)$-point designs even for small $N$.

\medskip\medskip\medskip

{\bf Acknowledgements.}
This work has been supported in part by the Collaborative
Research Center ``Statistical modeling of nonlinear dynamic processes'' (SFB 823, Teilprojekt C2) of the German Research Foundation (DFG).
The research of H. Dette reported in this publication was also partially supported by the National Institute of
General Medical Sciences of the National Institutes of Health under Award Number R01GM107639.
The content is solely the responsibility of the authors and does not necessarily
 represent the official views of the National
Institutes of Health. 
We would also like to thank Martina Stein who typed parts of this paper with considerable technical expertise.
The work of Andrey Pepelyshev was partly
supported by Russian Foundation of Basic Research,
project 12-01-00747.

\begin{appendix}

\section{Proof of main results}
\label{appB}
\def\theequation{A.\arabic{equation}}
\setcounter{equation}{0}

\subsection{Explicit form of the inverse of the covariance matrix of errors}

Here we state an auxiliary result, which
  gives an explicit form for
the inverse of the matrix $\mathbf{\Sigma} = (K(t_i,t_j))^N_{i,j=1}$, with a triangular covariance kernel $K$.
 We  did not find this result (as formulated below) in the literature.
 Versions of Lemma~\ref{lem:lemma1}, however, have been derived independently by different authors; see, for example, Lemma 7.3.2
in \cite{Z} and formula~(8) in \cite{Harman}. The proof follows from straightforward checking the condition
$\mathbf{\Sigma}^{-1}\mathbf{\Sigma}=\mathbf{\Sigma} \mathbf{\Sigma}^{-1} = \mathbf{I} $.

\begin{lemma}
\label{lem:lemma1}
{\it
Consider a symmetric  $N \times N $ matrix $\mathbf{\Sigma}=(\sigma_{i,j})_{i,j=1,\dots,N}$ which elements are defined by the formula
$ {\sigma}_{i,j}= u_i v_j$ for $1 \leq i \leq j \leq N$. Assume that $q_1<q_2< \ldots <q_N$ where  $q_i=u_i/v_i$.
Then the  inverse matrix $ \widetilde{\mathbf{\Sigma}}=\mathbf{\Sigma}^{-1}$ is a symmetric tri-diagonal matrix and its elements $\widetilde{{\sigma}}_{i,j}$ with $i \leq j$
   can be computed  as follows:
}
\bea
 \widetilde{{\sigma}}_{1,1}=
  {\frac { u_{2}}{u_1 v_1v_2 \left( q_2-q_1 \right) }}\, , \;\;\;  \widetilde{{\sigma}}_{N,N}=
 \frac {1}{ v_N^2 \left(  q_N  - q_{{N-1}} \right)  }\, ,
 \eea
\bea
 \widetilde{{\sigma}}_{i,i}=
\frac {q_{{i+1}}-q_{{i-1}}}
{v_i^2 (q_{{i}}-q_{i-1} ) (q_{{i+1}}- q_{i} ) }
\, \; ( i=2 , \ldots, N-1)\, ,
  \eea
\bea
 \widetilde{{\sigma}}_{i,i+1}=
-\frac{1}{ v_i v_{i+1}  ( q_{i+1}-q_{i} )}\, \; ( i=1 , \ldots, N-1)\, ,
  \eea
\bea
 \widetilde{{\sigma}}_{i,i+k}= 0 \, \; ( i=1 , \ldots, N-2,\;\; k \geq 2)\, .
  \eea
\end{lemma}

In our applications of Lemma~\ref{lem:lemma1} we assume that $\sigma_{i,j}=K(t_i,t_j)$ with the covariance kernel $K$ having the form \eqref{eq:cov_tr0}.

\subsection{Proof of Lemma~\ref{lem:2-1}}

Denote $K_{ij}=K(t_i,t_j)$, $f(t_i)=f_i$, $a_i=f_iw_i$, $i,j=1,\ldots,N$, $\mathbf{a}=(a_1,\ldots,a_N)^T$.
Then for any signed measure $\xi=\{t_1,\ldots,t_N;w_1,\ldots,w_N\}$
we have
\bea
D(\xi)=\frac{\sum_i\sum_j K_{ij} f_if_jw_iw_j}{(\sum_i f_i^2w_i)^2}=
\frac{\sum_i\sum_j K_{ij} a_ia_j}{(\sum_i f_ia_i)^2}
=\frac{\mathbf{a}^T\mathbf{\Sigma}\mathbf{a}}{(\mathbf{a}^T\mathbf{f})^2}.
\eea
Since $\mathbf{\Sigma}$ is symmetric and $\mathbf{\Sigma>0}$,
there exists $\mathbf{\Sigma}^{-1}$ and a symmetric matrix $\mathbf{\Sigma}^{1/2}>0$
such that $\mathbf{\Sigma}=\mathbf{\Sigma}^{1/2}\mathbf{\Sigma}^{1/2}$.
Denote $\mathbf{b}=\mathbf{\Sigma}^{1/2}\mathbf{a}$ and $\mathbf{d}=\mathbf{\Sigma}^{-1/2}\mathbf{f}$.
Then we can write the design optimality criterion $D(\xi)$ as $D(\xi)=\mathbf{b}^T\mathbf{b}/(\mathbf{b}^T\mathbf{d})^2$.
The Cauchy-Schwartz inequality gives for any two vectors $\mathbf{b}$ and $\mathbf{d}$ the inequality
$(\mathbf{b}^T\mathbf{d})^2\le(\mathbf{b}^T\mathbf{b})(\mathbf{d}^T\mathbf{d})$,
that is, $\mathbf{b}^T\mathbf{b}/(\mathbf{b}^T\mathbf{d})^2\geq 1/(\mathbf{d}^T\mathbf{d})$.
This inequality with $\mathbf{b}$ and $\mathbf{d}$ as above is equivalent to
$
 D(\xi)\geq {1}/{\mathbf{f}^T\,\mathbf{\Sigma}^{-1}\mathbf{f}}
$
for all $\xi$.
Equality  is attained if the vector $\mathbf{b}$ is proportional to the vector $\mathbf{d}$;
that is, if $b_i=cd_i$ for all $i$ and any $c\neq0$.
Finally, the equality $b_i=cd_i$ can be rewritten in the form
$w_i=c{(\mathbf{\Sigma}^{-1}\mathbf{f})_i}/{f(t_i)}$.

\subsection{Proof of Theorem \ref{th:as-opt-des-tr}}

Before starting the main proof we recall the definition of the design points \eqref{eq:des} and prove the following auxiliary
result.
\begin{lemma} \label{lemaux}
Assume that $q (\cdot)=u (\cdot) / v (\cdot)$ is a twice continuously differentiable function on the interval $[a,b]$. Then for all $i=1, \ldots, N-1$, we have
\be
\label{x m x1}
t_{i+1,N}-t_{i,N} &=&    \frac{1}{N Q^\prime (t_{i,N})} + O \Bigl(\frac1{N^2} \Bigr) \;\; {\rm as } \; N \to \infty.\\
\label{x m x}
\Delta_n = (t_{i+1,N}-t_{i-1,N})/2 &=&    \frac{1}{N Q^\prime (t_{i,N})}   \Bigl( 1+ O \Bigl(\frac1{N}\Bigr)\Bigr) \, \; {\rm as } \; N \to \infty.
\ee
\end{lemma}
{\bf Proof of Lemma \ref{lemaux}.}
Recall the definition $z_{i,N}=(i-1/2)/N$ ($i=1,\ldots , N$) and set
$$
m=q(a)= \min_{t \in [a,b]} q(t)~,~~  M=q(b)= \max_{t \in [a,b]} q(t).
$$
 From the definition of the function $Q$ in \eqref{qtraf}  we have
\be
\label{eq:q}
q(t_{i+1,N})- q(t_{i,N})= (M-m) ( z_{i+1,N}- z_{i,N} )= \frac{M-m}{N} \,
\ee
for all $i=1, \ldots, N-1$.  Observing   Taylor's formula yields  for any $z$
$$
Q^{-1} (z+\delta)= Q^{-1} (z)+ \delta \cdot { (Q^{-1} )}^\prime (z) + O(\delta^2) \;\; {\rm as } \; \delta \to 0.
$$
In this formula, set $z=z_{i,N}$ and   $\delta_N=1/N$ so that $z+\delta= z_{i+1,N}$. We thus obtain
$$
t_{i+1,N}-t_{i,N}= Q^{-1} (z_{i+1,N})= Q^{-1} (z_{i,N})+ \frac1{N} \cdot { (Q^{-1} )}^\prime (z_{i,N}) + O \Bigl(\frac1{N^2} \Bigr) \;\; {\rm as } \; N \to \infty.
$$
By using   \eqref{eq:des} and the relation
${\left(Q^{-1}\right)}^\prime (z)  = 1/Q^\prime (Q^{-1}(z))$
we can rewrite this  in the form \eqref{x m x1}.
The second statement  obviously follows from \eqref{x m x1}.

\medskip

{\bf Proof of Theorem \ref{th:as-opt-des-tr}.}  In view of Lemma \ref{lem:2-1} and \eqref{w10} - \eqref{w30}   we have
\bea
 w_{1,N} =
 \frac{ c_N u_{2}}{f_1 v_1 v_2 }
 \Bigl( \frac { f_{{1}}}{ u_{1}  } -  \frac{f_2}{u_2} \Bigr) \, \, ,\;\;\;
  w_{N,N}
 = \frac{c_N }{f_N v_N }
 \Bigl( \frac { f_{N}}{ v_{N}  } -  \frac{f_{N-1}}{v_{N-1}} \Bigr)  \, ,
   \eea
\bea
 w_{i,N}=
 \frac{c_N}{f_i v_i}
 \Bigl(
 \frac{2 f_{i}} {v_{i} }
 -
 \frac{f_{i-1}} {v_{i-1} }
-
 \frac{f_{i+1}} {v_{i+1}  }
    \Bigr) \, \;\; \mbox{ for $i=2, \ldots, N-1$},
  \eea
 where we have used the relations \eqref{eq:q}. Here $c_N$ is the normalization constant providing $\sum_{i=1}^N | w_{i,N}|=1$ and we use the notation  $u_i=u(t_{i,N})$, $v_i=(t_{i,N})$ and
 $f_i=f(t_{i,N})$.

Consider first $ w_{1,N}$. Denote $g (t)=f(t)/u(t)$, then
\be
\label{eq1}
 w_{1,N}/c_N =
 \frac{  u(t_{2,N})}{f(t_{1,N}) v(t_{1,N}) v(t_{2,N}) }
 \left( g (t_{1,N})- g (t_{2,N}) \right)\, ,
\ee
which gives
\be
\label{eq2}
 \frac{  u(t_{2,N})}{f(t_{1,N}) v(t_{1,N}) v(t_{2,N}) }=
 \frac{  u(a)}{f(a) v^2(a) } \Bigl( 1+ O \Bigl(\frac1{N}\Bigr)\Bigr)\, ,
\ee
\bea
g (t_{1,N})-g (a) = g ^\prime(a)  \left( t_{1,N}-a  \right)      +  O \left(   \left( t_{1,N}-a  \right)^2  \right)
=
g ^\prime(a) \cdot  \frac{1}{2N Q^\prime (a)} + O \Bigl(\frac1{N^2} \Bigr)
\eea
as $N \to \infty$. Similarly
\bea
g (t_{2,N})-g (a)
=
g ^\prime(a)   \frac{3}{2N Q^\prime (a)} + O \Bigl(\frac1{N^2} \Bigr)
\eea
yielding
\be
\label{eq3}
 g (t_{1,N})- g (t_{2,N})
=
-g ^\prime(a)  \frac{1}{N Q^\prime (a)} + O \Bigl(\frac1{N^2} \Bigr) \, .
\ee
Combining \eqref{eq1}, \eqref{eq2} and \eqref{eq3} we obtain
\be
\frac{w_{1,N}}{c_N}& =& \nonumber
-\frac1{N} \cdot  \frac{  u(a) g ^\prime(a)}{f(a) v^2(a) Q^\prime (a)} \Bigl( 1+ O \Bigl(\frac1{N}\Bigr)\Bigr)\\ &= &
\frac1{N} \cdot  \frac{  1 }{ v^2(a) Q^\prime (a)}  \Bigl[  \frac{u^\prime(a)}{u(a)} - \frac{f^\prime(a)}{f(a)}  \Bigr]
    \Bigl( 1+ O \Bigl(\frac1{N}\Bigr)\Bigr)
\label{eq4a}
\ee
as $N \to \infty$.
Similarly to \eqref{eq4a} we get the asymptotic expression for $ w_{N,N}$:
\be
\label{eq5}
\frac{w_{N,N}}{c_N} =
\frac1{N} \cdot  \frac{   h ^\prime(b)}{f(b) v(b) Q^\prime (b)}     \Bigl( 1+ O \Bigl(\frac1{N}\Bigr)\Bigr)
\ee
as $N \to \infty$.
Consider now the   weights
\be \label{eq w1}
 w_{i,N}=
 c_N \frac{
 2 h (t_{i,N})
 -
 h (t_{i-1,N})
-
 h (t_{i+1,N})
    }{f(t_{i,N}) v(t_{i,N})}
  \, \;\; \mbox{ ($i=2, \ldots, N-1$)}.
  \ee
Assume $N \to \infty$ and $i=i(N)$ is such that $i(N)/N= z + O(1/N)$ as $N \to \infty$ for some $z \in (0,1)$, and
set $t=Q^{-1} ( z )$.

We are going to prove that
\be \label{win1}
 \frac{w_{i,N}}{c_N}&=& \frac{
 2 h (t_{i,N})
 -
 h (t_{i-1,N})
-
 h (t_{i+1,N})
    }{f(t_{i,N}) v(t_{i,N})}
  \,
  \\
  \nonumber
  &=& \frac{1}{N^2 (Q^{\prime } (t))^2 f(t ) v( t )} \Bigl[ \frac{h ^{\prime }(t)  Q^{\prime \prime} (t) }{Q^{\prime } (t)} -h ^{\prime \prime}(t) \Bigr]   \Bigl( 1+ O \Bigl(\frac1{N}\Bigr)\Bigr)\, \\
     \label{win2}
  &=&- \frac{1}{N^2 Q^{\prime } (t) f(t ) v( t )} \Bigl[ \frac{h ^{\prime }(t) }{Q^{\prime } (t)} \Bigr]^\prime   \Bigl( 1+ O \Bigl(\frac1{N}\Bigr)\Bigr)\,.
     \ee

First, in view of \eqref{eq:des} we have
$t_{i,n}= t  + O \left(\frac1{N}\right)$ and hence
$$
f(t_{i,N}) v(t_{i,N})=  f(t ) v( t )  \Bigl( 1+ O \Bigl(\frac1{N}\Bigr)\Bigr) \;\;{\rm as}\; N \to \infty\, .
$$

Consider the numerator in \eqref{win1} and rewrite it as follows:
\bea
2 h (t_{i,N})
 -
 h (t_{i-1,N})
-
 h (t_{i+1,N})
 =  \left[2 h (\tilde{t}_{i,N})
 -
 h (t_{i-1,N})
-
 h (t_{i+1,N}) \right]+ 2 \left[h ({t}_{i,N}) - h (\tilde{t}_{i,N})
  \right]
\eea
where
$
\tilde{t}_{i,N}= (t_{i-1,N}+ t_{i+1,N})/2 \, .
$
We obviously have
$
 t_{i+1,N} =\tilde{t}_{i,N} + \Delta_N $ and $  t_{i-1,N}=  \tilde{t}_{i,N} - \Delta_N , $
where
$\Delta_N =(t_{i+1,N}- t_{i-1,N})/2$ is  defined in \eqref{x m x}. This yields
\be
\label{ll}
\frac{2 h (\tilde{t}_{i,N})
 -
 h (t_{i-1,N})
-
 h (t_{i+1,N}) } {\Delta_N^2}  = - h ^{\prime \prime}(t)   \Bigl( 1+ O \Bigl(\frac1{N}\Bigr)\Bigr) \, .
 \ee
Next we consider
\bea
 \frac{h ({t}_{i,N}) - h (\tilde{t}_{i,N})
  }{\Delta_N^2}=
  \frac{h ({t}_{i,N}) - h (\tilde{t}_{i,N})
  }{ {t}_{i,N}-\tilde{t}_{i,N}} \cdot \frac{ {t}_{i,N}-\tilde{t}_{i,N}}   {\Delta_N^2}.
  \eea
For the first factor we have
\bea
\frac{h ({t}_{i,N}) - h (\tilde{t}_{i,N})
  }{ {t}_{i,N}-\tilde{t}_{i,N}} = h ^{\prime }(t)   \Bigl( 1+ O \Bigl(\frac1{N}\Bigr)\Bigr) \, ,
  \eea
  while the second factor gives
 \bea
  \frac{ {t}_{i,N}-\tilde{t}_{i,N}}   {\Delta_N^2}&=&    2 \frac{ 2 t_{i,N}-t_{i+1,N}-t_{i-1,N}}   {(t_{i+1,N}- t_{i-1,N})^2} \\
  &=& 2 \frac{ 2 Q^{-1}  ( z_{i,N}  ) -Q^{-1}  ( z_{i+1,N}  )-Q^{-1}  ( z_{i-1,N}  )} {1/N^2}
  \Big(\frac{1/N}  { Q^{-1}  ( z_{i+1,N}  )- Q^{-1}  ( z_{i-1,N}  )} \Big)^2\\
   &=&-2 { (Q^{-1} )}^{\prime \prime}(z) /  (2 { (Q^{-1} )}^{\prime }(z)  )^2 \Big(  1+ O \Big( \frac1{N}\Big)\Big) =
    Q^{\prime \prime} (t) / (2 Q^\prime (t))
     \Big( 1+ O  \Big(\frac1{N} \Big)\Big)\, ,
     \eea
where we have used the relation ${\left(Q^{-1}\right)}^{\prime \prime} (z)=
-  Q^{\prime \prime} (z) / (Q^\prime (z))^3$
 in the last equation.
This gives, as $N \to \infty$,
\be
\label{llll}
 2\frac{h ({t}_{i,N}) - h (\tilde{t}_{i,N})
  }{\Delta_N^2}=
 2 \frac{h ({t}_{i,N}) - h (\tilde{t}_{i,N})
  }{ {t}_{i,N}-\tilde{t}_{i,N}} \cdot \frac{ {t}_{i,N}-\tilde{t}_{i,N}}   {\Delta_N^2}=  \frac{h ^{\prime }(t)  Q^{\prime \prime} (t)}{  Q^\prime (t)}
    \Bigl( 1+ O \Bigl(\frac1{N}\Bigr)\Bigr) .\;\;\;
  \ee

Combining the expressions
\eqref{x m x}, \eqref{win1}, \eqref{ll}  and \eqref{llll} yields the asymptotic expression  \eqref{win2} for ${w_{i,N}}/{c_N}$.

By noting that
$c_N=N C \left( 1+ O \left(\frac1{N}\right)\right)$ as $N \to \infty$ and that the asymptotic density of the points $t_{i,N}$ ($i=1, \ldots, N$) is  $Q^\prime(t)$ on the interval $[a,b]$,
we  deduce  the statement of the theorem as a consequence of  the asymptotic formulas \eqref{eq4a}, \eqref{eq5} and \eqref{win2}
for ${w_{1,N}}/{c_N}$, ${w_{N,N}}/{c_N}$ and ${w_{i,N}}/{c_N}$ respectively.

\subsection{Proof of Theorem \ref{th:opt-des-Wiener}}

By Theorem 3.3 in \cite{DetPZ2012} a design minimizes the functional \eqref{eq:Phi} if the identity
 \be
\label{blue_brown}
\int_{ a}^{ b} \min (s,t)   f(t) d   \xi(t) = \lambda    f(s)
\ee
holds $\xi\!-\!a.e.$, where $\lambda$ is some constant.
We consider the design $\xi= \xi ^*$  defined by \eqref{optdes} and \eqref{mass0a} and verify for this design the condition \eqref{blue_brown}.
  To do this we calculate by partial integration
\bea
\frac {1}{c} \int_{ a}^{ b} \min (s,t)   f(t)  p (t) dt &=& \int^s_{  a} t(-   f^{\prime \prime} (t))dt + \int^{ b}_s s (-   f^{\prime \prime} (t)) dt \\ \nonumber
 &=& - \Bigl \{ t   f^\prime (t) \Big |^s_{ a} - \int^s_{ a}   f^\prime (t) dt \Bigr \} - s \Bigl \{   f^\prime (  b) -   f^\prime (s) \Bigr \} \\ \nonumber
& =& ( a   f^\prime (  a) -  f(  a)) - s   f^\prime ( b) +   f(s).
\eea
Observing the definition of the masses in \eqref{mass0a}, the identify \eqref{blue_brown} follows with $\lambda=c$. This proves the first part of Theorem \ref{th:opt-des-Wiener}.

For a proof of the second statement consider a linear unbiased estimator $\hat \theta_{\mu^*}$ in model \eqref{onepar} based on the full trajectory, where
$\mu^*(dt)=f(t) \xi^*(dt)$ and $\xi^*$ is the design in \eqref{optdes}, \eqref{mass0a} with a constant $c$ chosen  such that $\hat \theta_{\mu^*}$ is unbiased, that is
$$
c^*= \Big[ \frac {f^2(a)}{a} + \int^b_a (f'(t))^2dt \Big]^{-1}.
$$
Standard arguments of optimal design theory show that $\mu^*$ minimizes $\Phi$ (that is, $\hat \theta_{\mu^*}$ is BLUE in model \eqref{onepar} where the full trajectory can be observed) if and only if the inequality
\be \label{neu2}
\Phi (\mu^*, \nu) = \int^b_a \int^b_a K(x,y) \mu^* (dx) \nu (dy) \geq \Phi (\mu^*)
\ee
holds for all signed measures $\nu$ satisfying $\int^b_a f(t) \nu (dt)=1$. Observing this condition and the identity \eqref{blue_brown} we obtain
$$
\Phi (\mu^*, \nu) = c^* \int^b_a f(s) \nu (ds) = \Bigl[ \frac {f^2(a)}{a} + \int^b_a (f^\prime(t))^2dt\Bigr]^{-1} = \Phi (\mu^*)
$$
for all signed measures $\nu$ on $[a,b]$ with $\int^b_a f(t) \nu (dt) =1$. By \eqref{neu2} $\mu^*$ minimizes $\Phi$. Consequently, the corresponding estimator $\hat \theta_{\mu^*}$ is BLUE with minimal variance
$$
D^* =c^* = \Bigl  [ \frac {f^2(a)}{a} + \int^b_a (f^\prime (t))^2 dt \Bigr ]^{-1}.
$$

\subsection{Proof of Theorem \ref{th:op-des-tr}}

Let $\{ \tilde \varepsilon (s) | s \in [\tilde a, \tilde b]$ be a Brownian motion on the interval $[\tilde a,\tilde b]$ and consider the regression model \eqref{onepar} with some function $\tilde f(s)$ and the error
process.
By Theorem \ref{th:opt-des-Wiener}
 the optimal design is given by
$$\tilde\xi^*(ds)=\tilde P_{\tilde a}\delta_{\tilde a}(ds)+\tilde P_{\tilde b}\delta_{\tilde b}(ds)+\tilde p(s)ds$$
with
$$
\tilde P_{\tilde a}=c \ \frac {\tilde f(a)-\tilde f^\prime (\tilde a) \tilde a}{\tilde a \tilde f(\tilde a)} \, ,\;\;
 \tilde P_{\tilde b}=c \ \frac {\tilde f^\prime (\tilde b)}{\tilde f(\tilde b)} \qquad {\rm{and}} \qquad
 \tilde p (s)= - c \frac {\tilde f ^{\prime \prime}(s)}{\tilde f(s)} \, .
$$
We shall now use Theorem \ref{th:mult-Phi=tPhit} to derive
the optimal design $\xi^*(dt)$ for the original  regression model \eqref{onepar} with regression function $f(t)$ and  covariance kernel $K(t,t')$ from
the design $\tilde\xi^*(ds)$ for the function $\tilde f(s)=h(q^{-1}(s))$, where $h(t)=f(t)/v(t)$.

For the Brownian motion, the covariance function is defined by \eqref{eq:cov2} with
$\tilde v(t)=1$ and $\tilde q(t)=t$ so that by  \eqref{eq:ab} we have $\beta(t)= q(t)$, $\alpha(t)= v(t)$ and $\tilde\alpha(t)=1/ v( q^{-1}(t)).$
According to \eqref{eq:mult-7a} the optimal design $d\tilde\xi^*(s)$ transforms
to $d  \xi^* (t) = \tilde  \alpha^2 ( \beta (t)) d \tilde{ \xi}^*( \beta (t) ) = d \tilde \xi^* (q(t))/\nu(t)$.

Consider first the mass at $b$. We have
$\tilde P_{\tilde b}=c {\tilde f^\prime (\tilde b)}/{\tilde f(\tilde b)}$. By using the transformation of $t$ into   $s= q^{-1}(t) $,
we obtain
$$
 P_{ b}= \frac {\tilde P_b}{v^2(b)}=c\frac {\tilde f^\prime (\tilde b)}{\tilde f(\tilde b)v^2(b)}
 =c \frac{ h^\prime ( b)}{q^\prime( b) v^2(b)  h(b)}
 =c \frac{ h^\prime ( b)}{q^\prime( b) v( b) f( b)}\, ,
$$
as required.
 From the representation of $\tilde P_{\tilde a}$ 
we obtain by similar arguments
$$
 P_{ a}=\frac {\tilde P_a}{v^2(a)}
 =c \  \frac { h( a)- a h^\prime ( a)/ q^\prime( a) }{ q^\prime( a) v^2( a)  h( a)}.
$$

Let us now consider the density $\tilde p(s),$ $s \in [\tilde a,\tilde b]$,
and rewrite $ d\tilde\xi^*_{p}( \beta (t) )$, the absolutely continuous part of the  measure $  \tilde\xi^*$.
The transformation of the variable $s$ into $t= q^{-1}(s) \in [a,b]$ induces the density
\be
\label{eq:dens1M}
d \tilde\xi^*_{p}( \beta (t) )=\tilde p(q(t)) q^\prime(t)= -c q^\prime(t)  \frac {\tilde f^{\prime \prime}(q(t))}{\tilde f(q(t))}\, .
\ee
Differentiating the equality $\tilde f(s)=h(q^{-1}(s))$, we have
\bea
\tilde f^{\prime \prime}(s)= \left(   h^{\prime }(q^{-1}(s)) \cdot (q^{-1}(s))^\prime   \right)^{\prime }=
    h^{\prime \prime} (q^{-1}(s)) \cdot \left(  (q^{-1}(s))^\prime     \right)^2 + h^{\prime} (q^{-1}(s)) \cdot  \left(q^{-1}(s) \right)^{\prime \prime}.
\eea
Now we obtain
 \bea
\tilde f^{\prime \prime}( q(t)) =
   \frac{ h^{\prime \prime} (t) }{  (q^{\prime}(t))^2}
      - h^{\prime} (t) \cdot  \frac{q^{\prime \prime} (t) } {(q^{\prime}(t))^3}
\eea
Inserting this into \eqref{eq:dens1M} and taking into account that $\tilde f (q(t)) = h(t)$, we obtain
the density
 \be \label{eq:dens2}
 d \tilde\xi^*_{p}( \beta (t) ) =c \, \frac{1} {h(t) q^{\prime}(t)} \Bigl(
  h^{\prime} (t) \cdot  \frac{q^{\prime \prime} (t) } {q^{\prime}(t)} -  h^{\prime \prime} (t) \Bigr)= - \frac{c} {h(t) }
  \left[ \frac{  h^{\prime} (t)  } {q^{\prime}(t)}  \right]^\prime \, .
\ee
In view of the relation
$d  \xi^* (t) = \tilde  \alpha^2 ( \beta (t)) d  \tilde\xi^*( \beta (t) )$
   we need to divide the right hand side in \eqref{eq:dens2} by $v^2(t)$ and obtain the expression for the density
\eqref{eqpA}. This completes the proof of Theorem \ref{th:op-des-tr}.

\section{Gaussian processes with triangular covariance kernels }
\label{appA}
\def\theequation{B.\arabic{equation}}
\setcounter{equation}{0}

\label{sec:Doob}

\subsection{Extended  Doob's representation}

Assume that $\{\ve(t)  | \; t \in [a,b]\}$  is a    Gaussian   process with  covariance kernel $K$ of the form \eqref{eq:cov_tr0}; that is,
$K(t,t')=u(t)v(t')$ for $t \leq t'$,
where $u(\cdot)$ and $v(\cdot)$ are    functions defined on the interval $[a,b]$.
According to the terminology introduced in  \cite{mehr1965certain} kernels of the form \eqref{eq:cov_tr0}
are  called {\it triangular}.
 An alternative way of writing these covariance kernels is
\be
\label{eq:cov_tr1}
\label{eq:cov1}
K(t,t')=v(t) v(t') \min\{q(t),q(t')\} \;\; {\rm for }\;\; t, t' \in [a,b]\,,
\ee
where  $q(t)= u(t)/v(t)$.  We assume that
$\ve(t)$ is non-degenerate
  on  the open interval $(a,b)$, which implies that
the function $q$ is strictly increasing and continuous on the interval $[a,b]$  [see \cite{mehr1965certain}, Remark 2].
Moreover, this function is also positive on the interval $(a,b)$ [see Remark 1 in \cite{mehr1965certain}], which
 yields that the functions  $u$ and $v$ must   have the same sign and can   be assumed to be  positive on the interval $(a,b)$ without loss of generality.
We repeatedly use  the following extension of the celebrated Doob's representation [see \cite{doob1949heuristic}], which relates to two Gaussian processes (on compact intervals) by a time-space transformation.


\begin{lemma}
\label{lem:tX-as-aXb}
 Let $\{\varepsilon(t) | \; t \in [a,b]\}$ be a non-degenerate Gaussian
   process   with zero mean and
  covariance function  \eqref{eq:cov1} and let $\tilde v$ and $\tilde q$
  be  continuous positive  functions on $[ \tilde a, \tilde b]$,
  such that $\tilde q$ is strictly increasing and $\tilde q([\tilde a, \tilde b])= q ([a,b])$. Define the transformations
  $\tilde \beta: [\tilde a, \tilde b] \to [ a,  b]$ and $\tilde \alpha: [\tilde a, \tilde b] \to \mathbb{R}_+ $ by
\be
 \label{eq:abT} \tilde \beta(s) = q^{-1} ( \tilde q(s) ) \, ,\;\;\;\tilde \alpha(s) = \tilde v(s) / v( \tilde \beta(s) )\, .
 \ee
Then the Gaussian   process $\{ \tilde \varepsilon (t) | \; t \in [\tilde a, \tilde b] \}$ defined by
\be
 \label{eq:2}
   \tilde \varepsilon(s)= \tilde \alpha(s) \varepsilon(\tilde \beta(s) )\,
\ee
has zero mean and  the covariance function is given by
 \be
 \label{eq:cov2}
    \tilde K(s,s')=\mathbb{E}[ \tilde \varepsilon(s) \tilde \varepsilon(s')] = \tilde v(s) \tilde v(s')   \min (\tilde q(s),\tilde q(s'))\, .
 \ee
 Conversely, the  Gaussian process   $ \varepsilon(t )$ can be expressed via $ \tilde \varepsilon(s )$
by the transformation
\be
 \label{eq:3}
    \varepsilon(t )=  \alpha(t) \tilde \varepsilon(\beta(t) )\, ,
\ee
where
\be
 \label{eq:ab}
 \beta(t)= \tilde q^{-1} (q(t))\, , \;\; \alpha(t) = v(t)/ \tilde v (\beta(t))\, .
\ee
\end{lemma}

\noindent
{\bf Proof.}   Since $\{\varepsilon(t)|t \in [a,b]\}$ is Gaussian and has zero mean,  the process defined by \eqref{eq:2} is also Gaussian and  has zero mean.
For the covariance function of the  process \eqref{eq:2}  we have
\bea
\mathbb{E} \left[ \tilde \varepsilon(s) \tilde \varepsilon(s' )\right]&=& \tilde \alpha(s) \tilde \alpha(s') \mathbb{E}  \left[ \varepsilon( \tilde \beta(s )) \varepsilon(\tilde \beta(s' ))\right]  \\
&=& \tilde \alpha(s) \tilde \alpha(s') v( \tilde \beta(s )) v(\tilde \beta(s' )) \min \left[ q( \tilde \beta(s )),\, q(\tilde \beta(s' ))\right]  \\
&=& {\tilde v(s)}{\tilde v(s')}  \min \left[  \tilde q(s ),\, \tilde q(s' )\right]  = \tilde K (s,s')\, .
\eea
The second part of the proof follows by the same arguments and the details are therefore omitted.
\hfill $\Box$

\medskip

\begin{remark}~  \label{rema1}  \rm

(a) The classical result of Doob  is a particular case of \eqref{eq:3} when $\tilde \varepsilon(t )=W(t)$ is the Brownian motion with
  covariance function $\tilde K (t,s)= \min (t,s) $. In this case we have $\tilde v (t)=1$, \mbox{$\tilde q(t)=t$},  $\alpha(t) = v(t)$ and   $ \beta(t)= q(t)$.
Specifically, the Doob's representation is given by $\varepsilon(t)=v(t) W(q(t))$ [see \cite{doob1949heuristic}].

(b)   Both functions $\beta:\; [a,b] \to [\tilde a, \tilde b] $ and $\tilde \beta: [\tilde a, \tilde b] \to [ a,  b]$ are
positive strictly increasing functions and are inverses of each other; that is,
 \be
 \label{eq:4}
    \beta(t )=   \tilde \beta^{-1}(t) \, ,\;\; \forall ~t \in [a,b]\, .
 \ee

(c) The functions $\alpha (\cdot)$  and $\tilde \alpha (\cdot)$ are positive and satisfy  the relation
  \be
 \label{eq:6}
  \alpha(t) \cdot \tilde \alpha(\beta(t)) = \frac{ v(t)}{ \tilde v (\beta(t))} \cdot \frac{\tilde v(\beta(t)) }{ v( \tilde \beta( \beta(t)) )} = 1 \, ,\;\; \forall ~t \in [a,b].
 \ee

 (d) The properties (b) and (c)  imply that the transformation $\tilde \varepsilon\to \varepsilon$ defined by \eqref{eq:3} is the inverse of the transformation
$\varepsilon \to \tilde \varepsilon$ defined in \eqref{eq:2}.
\end{remark}

\subsection{Transformation of regression models}
\label{sec:trans}

Associated with the transformation of the  triangular covariance kernels
there exists a canonical transformation for the corresponding regression models.
To be precise, consider the regression model \eqref{eq:model} or its continuous time version \eqref{mod1cont},
where the  covariance kernel $K(\cdot,\cdot)$ has the form \eqref{eq:cov1}. Recall the definition of
  the transformation $\beta:\; [a,b] \to [\tilde a, \tilde b] $ defined in \eqref{eq:ab}, which maps the observation points $t_j$ to  $\tilde t_j=\beta(t_j)$,  $j=1, \ldots, N$ and define
\be
 \label{eq:9a}
 \tilde f(s)  =  \frac{f( \tilde \beta (s))}{\alpha (\tilde \beta (s))},
 ~\tilde\varepsilon(s)=
 \frac{\varepsilon ( \tilde \beta (s))}{\alpha (\tilde \beta (s))},
 ~\tilde y(\tilde t_j)=\frac{y(t_j)}{\alpha (t_j)}\, ,
\ee
where $s \in [\tilde a, \tilde b]$ so that $ \tilde \beta (s) \in [a,b]$.
The regression model \eqref{eq:model} can now be rewritten in the form
\be
\label{eq:modela}
 \tilde y(\tilde t_j)= \theta^T \tilde f(\tilde t_j)+\tilde \varepsilon(\tilde t_j),~ \tilde t_j\in[\tilde a,\tilde b],
 ~ j=1, \ldots, N.
\ee
The errors $\tilde\varepsilon(\tilde t_j)$ in \eqref{eq:modela} have zero mean and, by Lemma~\ref{lem:tX-as-aXb} and the identity \eqref{eq:6},  their covariances are given by
\be
\label{eq:modelaCOV}
~\mathbb{E}[\tilde\varepsilon(\tilde t_i) \tilde\varepsilon(\tilde t_j)]=
\tilde K(\tilde t_i,\tilde t_j).
\ee

Hence we have transformed the regression observation scheme \eqref{eq:model} with error covariances $\mathbb{E}[\varepsilon(t_i)\varepsilon(t_j)]=K(t_i,t_j)$ to the scheme
\eqref{eq:modela} with covariances \eqref{eq:modelaCOV}.
Conversely, we can transform the model \eqref{eq:modela} with covariances \eqref{eq:modelaCOV} to the model \eqref{eq:model} using the  transformations
\be
 \label{eq:8a}
f(t) = \frac{\tilde  f( \beta (t))}{\tilde  \alpha ( \beta (t))} \, , \;\; \varepsilon(t) = \frac{\tilde  \varepsilon( \beta (t))}{\tilde  \alpha ( \beta (t))} \, , \; t \in [a,b] \, .
\ee

\begin{lemma}
\label{lem:transf-models}
The  transformation $f \to \tilde f$  defined in \eqref{eq:9a} is an
 inverse  to the transformation $\tilde f \to  f $ defined in  \eqref{eq:8a}.
\end{lemma}

\noindent
{\bf Proof.}
Inserting the expression for $\tilde f$ from \eqref{eq:9a} into \eqref{eq:8a}, we have
\bea
 f(t)= \frac{\tilde  f( \beta (t))}{\tilde  \alpha ( \beta (t))}=  \frac{f( \tilde \beta (\beta (t)))}{\alpha (\tilde \beta (\beta (t))) \, \tilde  \alpha ( \beta (t))} = \frac{f(t)}{\alpha (t) \, \tilde  \alpha ( \beta (t))}= f(t),
\eea
where we have used  the identities $ \tilde \beta (\beta (t))= t$, see \eqref{eq:4}, and $\alpha (t) \, \tilde  \alpha ( \beta (t))=1$, see
  \eqref{eq:6}.

\medskip

\subsection{Transformation of designs}

In this section we consider a transformation of the matrix-weighted designs
under a given transformation of the regression models. In the one-parameter case with $m=1$, these matrix-weighted designs become signed measures; that is,
 signed designs as considered in Section \ref{sec4}.
In this section, it is convenient to define all integrals  as Lebesgue-Stieltjes integrals with respect to the distribution functions of the measures $\zeta$ and $\tilde \zeta$.

To be precise, let $d\xi(t)=\mathbf{O}_\xi(t)d \zeta(t) $ be a matrix-weighted design on the interval  $t \in [a,b]$.
Recalling the definition of $\alpha, \tilde \alpha$ and $\beta,
\tilde \beta$ in \eqref{eq:abT} and \eqref{eq:ab}
we define  a matrix-weighted design $d \tilde \xi(s)= \tilde{\mathbf{O}}_{\tilde \xi} (s) d \tilde \zeta(s)$ by
\be
 \label{eq:mult-7}
  d \tilde \zeta (s) =  d \zeta(\tilde \beta (s) )\, \;\; {\rm and} \;\;  \tilde{\mathbf{O}}_{\tilde \xi} (s) = \alpha^2 (\tilde \beta (s)) \mathbf{O}_\xi(\tilde \beta (s) )\, .
\ee
Note that $\tilde \zeta$ and $\zeta$ are probability measures on the intervals $[\tilde a, \tilde b]$ and $[a,b]$, respectively.
Similarly, for a given  matrix-weighted design
$d \tilde \xi(s)= \tilde{\mathbf{O}}_{\tilde \xi} (s) d \tilde \zeta(s)$  on the interval $[ \tilde a,  \tilde b] $
we define  a matrix-weighted design $d\xi(t)=\mathbf{O}_\xi(t)d \zeta(t) $ on the interval $[a,b]$ by
\be
 \label{eq:mult-7a}
 d  \zeta (t) =  d \tilde  \zeta( \beta (t) ) \, \;\; {\rm and} \;\;
 \mathbf{O}_\xi (t) = \tilde  \alpha^2 ( \beta (t)) \tilde{\mathbf{O}}_{ \tilde  \xi}( \beta (t) ).
\ee
Similar to Lemma \ref{lem:transf-models} we can see
that the  transformation $\tilde \xi \to  \xi$ defined by \eqref{eq:mult-7a} is the
 inverse  to the transformation $\xi \to \tilde \xi$ defined by  \eqref{eq:mult-7}.

\medskip

For the following discussion we  recall the definition   of the covariance matrix
$\mathbf{D} (\xi)$ in \eqref{D_Matrix}.
For the model \eqref{eq:modela},
the covariance matrix of   the  design $d \tilde \xi(s)= \tilde{\mathbf{O}}_{\tilde \xi} (s) d \tilde \zeta(s)$,  defined by \eqref{eq:mult-7}, is given by
\be
\label{D_Matrix1}
\tilde{\mathbf{D}} (\tilde \xi)= \tilde{\mathbf{M}}^{-1}(\tilde\xi) \tilde{\mathbf{B}}(\tilde\xi) (\tilde{\mathbf{M}}^{-1}(\tilde\xi))^T,
\ee
where
{
\bea
\tilde{\mathbf{B}}(\tilde\xi) =  \int_{\tilde a}^{\tilde b} \int_{\tilde a}^{\tilde b}  \tilde K(t,s)  {\tilde{\mathbf{O}}}_{\tilde \xi}(t)\tilde f(t)
 ({\tilde{\mathbf{O}}}_{\tilde \xi}(s)\tilde f(s))^T
d \tilde\zeta (t) d \tilde\zeta (s), \;\;
\tilde{\mathbf{M}}(\tilde\xi)= \int_{\tilde a}^{\tilde b} {\tilde{\mathbf{O}}}_{\tilde \xi}(t) {\tilde f}(t) {\tilde f}^T(t) d \tilde\zeta (t)\,
\eea}
and the kernel $\tilde K$ is defined by \eqref{eq:cov2}.


\begin{theorem}
\label{th:mult-Phi=tPhit}
 For any matrix-weighted design $d\xi(t)=\mathbf{O}_\xi(t)d \zeta(t) $ and the corresponding matrix-weighted design $\tilde \xi$ defined by \eqref{eq:mult-7},  we have
$\mathbf{D} (\xi) = \tilde{\mathbf{D}} (\tilde \xi).$ In particular, $\mathbf{D}^*=\tilde{\mathbf{D}}^*$, where $\mathbf{D}^*$ and $\tilde{\mathbf{D}}^*$
are the covariance matrices of the BLUE in the continuous time models \eqref{mod1cont} and in the model $\{\theta^T \tilde f(s) + \tilde \varepsilon(s) | s \in [\tilde a, \tilde b]\}$, respectively.
\end{theorem}

\noindent
{\bf Proof.}     Using the variable transformation  $\tilde \beta (s)=t$ and \eqref{eq:9a}, we have
\bea
 \tilde{\mathbf{M}}(\tilde \xi)&=& \int {\tilde{ \mathbf{O}}}_{\tilde \xi}(s) {\tilde f}(s) {\tilde f}^T(s)  d \tilde\zeta (s)=
 \int \frac{\mathbf{O}_\xi(\tilde \beta (s)) f( \tilde \beta (s))}{\alpha (\tilde \beta (s))}
 \frac{f^T( \tilde \beta (s))}{\alpha (\tilde \beta (s))} \cdot \alpha^2 (\tilde \beta (s)) d \zeta(\tilde \beta (s) )\\
 &=&
 \int \mathbf{O}_\xi(t) f(t) f^T(t) d \zeta (t) = \mathbf{M}(\xi).
\eea
 Next, we calculate the corresponding expression for $ \tilde{\mathbf{B}}(\tilde \xi)$, that is
 \bea
\tilde{\mathbf{B}}(\tilde \xi)& =&  \int \int  \tilde K(x,y)
 \tilde{\mathbf{O}}_{\tilde \xi}(x) \tilde f(x)  (\tilde{\mathbf{O}}_{\tilde \xi}(y)\tilde f(y))^T d \tilde\zeta (x) d \tilde\zeta (y)\\
&=&
\int \int \tilde v(x) \tilde v(y)   \min (\tilde q(x),\tilde q(y))
 \tilde{\mathbf{O}}_{\tilde \xi}(x) \tilde f(x)  (\tilde{\mathbf{O}}_{\tilde \xi}(y)\tilde f(y))^T d \tilde\zeta (x) d \tilde\zeta (y)\\
&=&
\int \int \tilde v(x) \tilde v(y)   \min (\tilde q(x),\tilde q(y)) \frac{\mathbf{O}_\xi(\tilde \beta (x))f( \tilde \beta (x))}{\alpha (\tilde \beta (x))}
\frac{(\mathbf{O}_\xi(\tilde \beta (y))f( \tilde \beta (y)))^T}{\alpha (\tilde \beta (y))} \times\\
&& \alpha^2 (\tilde \beta (x)) d \zeta(\tilde \beta (x) )  \alpha^2 (\tilde \beta (y)) d \zeta(\tilde \beta (y) )
\eea
Define $s=\tilde \beta (x)$ and $t=\tilde \beta (y)$ so that $x= \tilde \beta ^{-1} (s) = \beta(s)$ and similarly $y=  \beta(t)$.
Changing the variables in the integrals above we obtain
{\small
\bea
\tilde B(\tilde \xi)& \!\!=\!\!&
\int \!\int \tilde v(\beta(s)) \tilde v(\beta(t))   \min (\tilde q(\beta(s)),\tilde q(\beta(t)))
\mathbf{O}_\xi(s)f( s) (\mathbf{O}_\xi(t)f(t))^T    \alpha (s) \alpha(t) d \zeta(s) d\zeta(t).
\eea}
Using the definition of  $\beta$ in \eqref{eq:ab} yields
$
\tilde q(\beta(t))=\tilde q(\tilde q^{-1} (q(t))) =q(t)
$
and by
the definition of  $\alpha$  in \eqref{eq:ab}  we finally  get
{\small
 \bea
\tilde{\mathbf{B}}(\tilde \xi)& =&
\int \!\int \tilde v(\beta(s)) \tilde v(\beta(t))   \min ( q(s), q(t)) \mathbf{O}_\xi(s)f( s) (\mathbf{O}_\xi(t)f(t))^T
\frac{v(s)} {\tilde v (\beta(s))}  \frac{v(t)} {\tilde v (\beta(t))}   d \zeta(s) d\zeta(t)\\
& =&
\int \!\int    \min ( q(s), q(t)) \mathbf{O}_\xi(s)f( s) (\mathbf{O}_\xi(t)f(t))^T v(s)  v(t)  d \zeta(s) d\zeta(t) = \mathbf{B}( \xi) \, .
\eea}
The result $\mathbf{D} (\xi) = \tilde{\mathbf{D}} (\tilde \xi)$  follows now from the definitions \eqref{D_Matrix} and \eqref{D_Matrix1}.

\end{appendix}

\setlength{\bibsep}{1pt}
\begin{small}
\bibliography{opt_signed_designsa}
\end{small}

\end{document}